\documentclass[aps,showpacs,preprintnumbers,amsmath,amssymb]{revtex4}

\usepackage{graphicx}
\usepackage{dcolumn}
\usepackage{bm}

\begin{document}
\preprint{APS/123-QED}

\title{$N$-Point Vertex Functions, Ward-Takahashi Identities and \\
Dyson-Schwinger Equations in Thermal QCD/QED\\
in the Real Time Hard-Thermal-Loop Approximation}

\author{Yuko FUEKI}%
 \email{yfueki@phys.nara-wu.ac.jp}
 \affiliation{Department of Physics,  Nara Women's University,
   Kita Uoya-nishimachi,Nara 630-8506, Japan}%

\author{Hisao NAKKAGAWA}%
 \email{nakk@daibutsu.nara-u.ac.jp}
\author{Hiroshi YOKOTA}%
 \email{yokotah@daibutsu.nara-u.ac.jp}
\author{Koji  YOSHIDA}%
 \email{yoshidak@daibutsu.nara-u.ac.jp}
\affiliation{Institute for  Natural Sciences,  Nara University,
1500 Misasagi-cho,
 Nara 631-8502, Japan}

\date{\today}

\begin{abstract}
In this paper we calculated the $n$-point hard-thermal-loop (HTL) vertex
 functions in QCD/QED for  $n$= 2, 3 and 4 in the physical
 representation in the real time formalism (RTF). The result showed that
 the $n$-point HTL vertex functions can be classified into two groups,
 a) those with odd numbers of external retarded indices,  and b) the
 others with even numbers of external  retarded indices.  The $n$-point
 HTL vertex functions with one retarded index, which obviously belong to
 the first group a), are nothing but the HTL vertex functions that
 appear in the imaginary time formalism (ITF), and vise versa. All the HTL
 vertex functions belonging to the first group a) are of  $O(g^2T^2)$ , and
 satisfy among them the simple QED-type Ward-Takahashi identities, as
 in the ITF. Those vertex functions belonging to the second group b)
 never appear in the ITF, namely their existence is characteristic of 
 the RTF,  and their HTL's have the high temperature behavior of 
 $O(g^2T^3)$, one-power of $T$ higher than usual. Despite this 
 difference we could verify that those HTL vertex functions belonging
 to the second group b) also satisfy among themselves the QED-type 
 Ward-Takahashi identities, thus guaranteeing the gauge invariance of 
 the HTL's in the real time thermal  QCD/QED.
\end{abstract}

\maketitle

\makeatletter 
\@addtoreset{equation}{section}
\def\theequation{\thesection.\arabic{equation}}
\section{ Introduction and summary}

The hard-thermal-loop (HTL) resummed effective perturbation theory of
Braaten and Pisarski \cite{1,2} has given us in finite temperature (or 
thermal) field theory \cite{3,4} a procedure correctly 
estimating/extracting the dominant temperature effect due to the 
semi-classical thermal fluctuation.  HTL's are originally determined 
in the imaginary time formulation of thermal field theory \cite{1,2} 
by computing the one-loop diagrams. For calculating the Feynman diagrams
in thermal field theory there are essentially two different
formulations, namely the imaginary time formalism (ITF) and the real
time formalisn(RTF) \cite{3,4,5}. The ITF is restricted to calculating
the static quantities in equilibrium, while the RTF is indispensable for
analysing the dynamical quantities and for investigating the
out-of-(or non)-equilibrium system thus for studying the physics of 
quark-gluon plasma.

The relation between quantities calculated in the ITF and in the RTF has
been studied so far \cite{6}; 
the $n$-point function in the ITF corresponds to a specific sum of the
$n$-point functions in the RTF, the same is true for the $n$-point HTL
functions. In the ITF all the $n$-point HTL functions have been obtained,
and are shown to be ultraviolet finite, gauge invariant and to satisfy
simple Ward-Takahashi identities \cite{1,2}. Also in the RTF the 
$n$-point HTL functions ($n$=2, 3, 4) and their spectral
functions have been partly determined \cite{7,8}, but the explicit 
expressions of the full  $n$-point HTL functions have not been 
consistently given yet \cite{f1}. Even with such a 
limited knowledge on the $n$-point 
HTL functions in the RTF, analyses of the 3-point functions have shown 
\cite{8,9} that there exist the HTL functions being proportional to 
$g^2T^3$ ($g$: the coupling constant, and $T$: the temperature of the 
environment), never existing in the ITF where all the $n$-point HTL 
functions being proportional to $g^2T^2$. Here it is worth noting that 
in the RTF the Ward-Takahashi  identities have been confirmed to hold 
only among the ITF counterparts, namely among the special linear 
combinations of $n$-point HTL functions corresponding to those in the 
ITF, having not been seen anything among those HTL functions not 
existing in the ITF.        

Recently beginning of the relativistic heavy ion collision experiments 
at BNL-RHIC has attracted an increasing interest in studying the
physics in thermal QCD. Progress of the physics of quark-gluon plasma 
and the development of the calculational framework towards the system 
in non-(or out-of-)equilibrium have urged us to determine explicitly 
all the HTL's in the RTF with the relations such as the Ward-Takahashi 
identities they might have satisfied. 

In this paper we determine explicitly in the RTF, especially in the
"physical representation" in the closed-time-path (CTP), or the Keldysh
formalism \cite{4,5}, the HTL contributions to the $n$-point vertex 
functions in thermal QCD/QED with $n$=2, 3, 4  which are in urgent 
demand. Also studied are the structure of the Ward-Takahashi 
identities that might be satisfied among them.  As an application we 
write down the Dyson-Schwinger equations in the HTL approximation for 
the fermion mass function. Results can be summarized briefly as
follows \cite{f2};

i) In the RTF there are two types of $n$-point vertex functions having
the HTL contributions; one is the vertex function with $n$ external gauge
bosons ($n$g vertex function) and the other with a pair of
external fermions and ($n$-2) external gauge bosons (2f-($n$-2)g vertex
function). There are no other $n$-point HTL functions, thus there are no
$n$-point HTL functions with external ghost lines. This situation is
completely the same as in the ITF \cite{1}. 

ii) All the $n$-point HTL vertex functions with one retarded index in the
physical representation in the CTP formalism take exactly the same
expressions (thus are proportional to $g^2T^2$) as those analytically
continued  from the $n$-point HTL vertex functions determined in the ITF
through simple but corresponding continuation paths. They satisfy the
same Ward-Takahashi identities as those in the ITF \cite{1,2}. The 
$n$-point vertex functions with one retarded index ($V_{RAA}, V_{ARA}, 
V_{RAAA}, V_{ARAA}$, etc) are nothing but the functions exactly 
corresponding to the $n$-point vertex functions that can be calculated 
in the ITF, and are given by the specific sum (for details, see
section \ref{s2}) of those in the "single time representation" in the 
CTP formalism, where a given $n$-point Feynman diagram contains a 
Keldish index at each end (external vertex) taking values \{1,2\}, 
corresponding to the two branches of the closed-time-path contour. 

iii) The $n$-point ($n$=2, 3, 4) vertex functions with two retarded
indices ($V_{RRA}, V_{RRAA}$, etc) are characteristic of the RTF, 
never appearing in the ITF. The corresponding $n$-point HTL vertex 
functions must be classified into two groups: the $n$g vertex 
functions and the 2f-($n$-2)g vertex functions. Other types of 
$n$-point vertex functions in general never have the HTL 
contributions as mentioned in i) above. It is quite remarkable that 
\textit{in QCD the $n$g vertex functions with two retarded indices 
are proportional to $g^2T^3$} in contrast to the  $O(g^2T^2)$ 
behavior of the usual HTL vertex functions, thus are expected to play 
important roles in studying the temperature effects. Nevertheless we 
can verify that they satisfy the QED-type Ward-Takahashi identities 
between the corresponding $n$g- and ($n$-1)g HTL vertex functions, 
still guaranteeing the gauge invariance of the HTL approximation. Thus
we can prove, through the explicit calculations of the $n$-point
HTL vertex functions, the gauge invariance of the real time thermal 
QCD/QED in the HTL approximation.  It is also worth noting that 
\textit{the HTL contributions to the 2f-($n$-2)g vertex functions 
with two retarded indices totally vanish} \cite{f3}, thus in QED 
there are no additional HTL functions other than those appeared in the ITF.

iv) In performing the calculation of the Feynman diagrams in the RTF we
should be very careful for the treatment of the singular functions (the
Dirac-$\delta$ function and the principal part) appearing in the free
propagators. We must use the properly regularized forms during the
calculation and should take the limit $\varepsilon\ \rightarrow 0$ at the
end of all calculations, as noted by Landsman and van Wheert
\cite{4}. 

v) We explicitly write down the HTL resummed Dyson-Schwinger equation 
for the physical fermion mass function, namely the retarded fermion 
self-energy function, in thermal QCD/QED, which can be used to 
investigate the nature of the chiral phase transition at finite 
temperature. Some comment on the double counting problem is also given.

\vspace{0.3cm}
This paper is organized as follows. In the next section \ref{s2} we
give a brief review of the "physical representation" in the Keldysh
or the CPT formalism. In section \ref{s3} the $n$-point HTL vertex 
functions ($n$=2, 3, 4) are explicitly determined in
the physical representation. The necessity of the use of regularized
form of the singular functions is demonstrated. The HTL Ward-Takahashi
identities between the four- and three-point functions together with the
three- and two-point functions are explicitly verified. As an
application the Dyson-Schwinger equations in the HTL approximation for
the physical fermion mass function $\Sigma_R$ is derived in section
\ref{s5}. Conclusions and some discussion are given in the last section
\ref{s6}.

\section{Real Time Closed-Time-Path or the Keldysh Formalism}
\label{s2}

We use the closed-time-path (CTP) or the Keldysh formalism \cite{4,5}
of the RTF throughout this paper. In this formalism there are two 
familiar representations, namely by following the terminology of 
Ref.\cite{4},  the "single time" representation and the "physical", 
or the "retarded-advanced" representation. In the single time 
representation the single-particle propagator for free bosons has 
the 2 $\times$ 2 matrix form

\begin{equation}
\label{2.1}
   \hat{D}(K) = \left( \begin{array}{ll}
                       D_{11}(K) & D_{12}(K) \\
                       D_{21}(K) & D_{22}(K) \\
                       \end{array}\right),
\end{equation}
$D_{ij} \ (i,j=1,2)$ are given in momentum space by
\begin{subequations}
\begin{eqnarray}
D_{11}(K) & = &  \frac{1+n_B(k_0)}{K^2-m^2+i\varepsilon} 
              - \frac{n_B(k_0)}{K^2-m^2-i\varepsilon}\ ,\\
D_{12}(K) & = &  [\theta(-k_0)+n_B(k_0)] \left(\frac 1{K^2-m^2+i\varepsilon} 
   - \frac 1{K^2-m^2-i\varepsilon}\right)\ , \\ 
D_{21}(K) & = &  [\theta(k_0)+n_B(k_0)] \left(\frac 1{K^2-m^2+i\varepsilon} 
             - \frac 1{K^2-m^2-i\varepsilon}\right)\ ,\\
D_{22}(K) & = &  \frac{n_B(k_0)}{K^2-m^2+i\varepsilon} 
                - \frac{1+n_B(k_0)}{K^2-m^2-i\varepsilon}\ ,
\end{eqnarray}
\end{subequations}
where $ K = (k_0,
\mbox{\boldmath{$k$}})$, $\theta$ denotes
the step function, and the equilibrium distribution function is given by
$ n_B(k_0) = 1/[\exp(\vert k_0\vert/T) -1]$. For fermions the bare 
propagator can be also written as the 2 $\times$ 2 matrix form

\begin{equation}
\label{2.3}
\hat{S}(K) =  \left( \begin{array}{ll}
                       S_{11}(K) & S_{12}(K) \\
                       S_{21}(K) & S_{22}(K) \\
                       \end{array}\right),
\end{equation}
and 
\begin{subequations}
\begin{eqnarray}
\label{2.4a}
S_{ij} & = &  (\not{\!K} -m)\overline{D}_{ij}(K)\ , \\
\label{2.4b}
\overline{D}_{ij} & = &  D_{ij}\{n_B(k_0) \rightarrow -n_F(k_0)\}\ 
,\  (i,j=1,2),
\end{eqnarray}
\end{subequations}
where the Fermi-Dirac distribution is given by $n_F(k_0) = 
1/[\exp(\vert k_0 \vert/T) +1]$. Components of these propagators are 
not independent of each other. There is an algebraic identity
\begin{equation}
\label{2.5}
G_{11} +G_{22} = G_{12} + G_{21},
\end{equation}
where $G$ stands for $D$ or $S$, respectively.

\vspace{0.3cm}
By an orthogonal transformation
\begin{eqnarray}
 \hat{G} & = &  Q^{-1} \tilde{G} Q,\  \tilde{G} = Q \hat{G}  Q^{-1};\
  G = D \ \mbox{or}\  S\ ,\\
Q & = & \frac1{\sqrt{2}}(1 - \sigma_2)\ ,
\end{eqnarray}
we arrive at the propagator in the physical, or the retarded-advanced 
representation,

\begin{equation}
\label{2.8}
\tilde{D}(K) = \left( \begin{array}{ll}
                       D_{AA}(K) & D_{AR}(K) \\
                       D_{RA}(K) & D_{RR}(K) \\
                       \end{array}\right),
\end{equation}
and $D_{\alpha\beta} \ (\alpha,\ \beta= A,R)$ are
\begin{subequations}
\begin{eqnarray}
\label{2.9a}  D_{AA}(K) & = & 0, \\
\label{2.9b}
D_{AR}(K) \ (\ \equiv  D_A(K)) & = &
\frac1{K^2 -  m^2  - i \mbox{sgn}(k_0)\varepsilon}\ ,\\
\label{2.9c}
D_{RA}(K) \ (\ \equiv D_R(K)) &  = &
\frac1{K^2 -  m^2  +i \mbox{sgn}(k_0)\varepsilon}\ ,\\
D_{RR}(K) \ (\ \equiv D_C(K)) &  =  &
\label{2.9d}(1+2n_B(k_0))\mbox{sgn}(k_0)[D_{RA}(K) - D_{AR}(K)]\ ,
\end{eqnarray}
\end{subequations}
for bosons and 
\begin{equation}
\label{2.10}
\tilde{S}(K) = \left( \begin{array}{ll}
                       S_{AA}(K) & S_{AR}(K) \\
                       S_{RA}(K) & S_{RR}(K) \\
                       \end{array}\right),
\end{equation}
\begin{subequations}
\begin{eqnarray}
\label{2.11a}  S_{AA}(K) & = &  0,\\
\label{2.11b}
S_{AR}(K) \ (\ \equiv 
S_A(K))  & = &  (\not{\!K} -m) 
\frac1{K^2 -  m^2  +i \mbox{sgn}(k_0)\varepsilon}\ ,\\
\label{2.11c}
S_{RA}(K) \ (\ \equiv
S_R(K)) & = &  (\not{\!K} -m) 
\frac1{K^2 -  m^2  -i \mbox{sgn}(k_0)\varepsilon}\ ,\\
\label{2.11d}S_{RR}(K) \ (\ \equiv 
S_C(K)) &  = &  (1- 2n_F(k_0))\mbox{sgn}(k_0)[S_{RA}(K) - S_{AR}(K)]\ ,
\end{eqnarray}
\end{subequations}
for fermions, where the last
equations in Eqs. (II.9) and (II.11) are consequences of the 
fluctuation-dissipation theorem.

Propagator (connected 2-point Green function) $G$ and the 1-particle 
irreducible 2-point vertex function $\Sigma$ satisfies the Dyson 
equation ($G=D$ or $S$, and correspondingly $\Sigma$ denotes the 
bosonic or fermionic 2-point vertex, i.e., self-energy function)
\begin{subequations}
\begin{eqnarray}
\label{2.12a}
\int \hat{\Sigma}  \sigma_3  \hat{G}  & = &   \sigma_3,  \ \ \ \
                     \int \hat{G} \sigma_3  \hat{\Sigma}  =  \sigma_3,\\
\label{2.12b}
 \int \tilde{\Sigma} \sigma_1  \tilde{G} & = &   \sigma_1,  \ \ \ \ 
                     \int \tilde{G} \sigma_1 \tilde{\Sigma}  =  \sigma_1.
\end{eqnarray}
\end{subequations}
In the above, the relations (\ref{2.1}), (\ref{2.3}),
 (\ref{2.5})-(\ref{2.8}), (\ref{2.10}) and (II.12) also hold for 
full propagators and full 2-point vertex functions. 

It is also worth noticing here that in the above equations for the
free propagators, Eqs. (\ref{2.1})-(II.4) and (\ref{2.8})-(II.11), we 
do not use the expressions that contain explicitly the singular
functions themselves, but rather use the functions with the 
regularization parameter $\varepsilon$, e.g.,
\begin{equation}
\label{2.13}
\frac{1}{[K^2 - m^2 + i \varepsilon]} = \mbox{PP}\frac{1}{[K^2 - m^2]}
 - i \pi \delta (K^2 - m^2),
\end{equation}
where
\begin{subequations}
\begin{eqnarray}
\label{2.14a}
\mbox{PP}\frac{1}{x} & \equiv &  \frac{x}{x^2 + {\varepsilon}^2},\\
\label{2.14b}
\pi \delta (x) & \equiv &  \frac{\varepsilon}{x^2 + {\varepsilon}^2}. 
\end{eqnarray}
\end{subequations}
Keeping $\varepsilon$ finite (i.e., $\varepsilon \neq 0$) till the end 
of all calculations then taking the limit  $\varepsilon \rightarrow 0$  
at the end is extremely important as we shall explicitly see in the
next section.

\section{The $n$-point ($n$=2, 3, 4) vertex functions in QCD/QED in
the HTL approximation}
\label{s3}

In this section we calculate the HTL contributions to the one particle 
irreducible $n$-point vertex functions ($n$=2, 3, 4) in the physical
or the retarded-advanced representation.  One can construct in general 
the $n$-point vertex functions in the physical or the
retarded-advanced representation from the components of the real-time 
$n$-point functions in the single time representation as in literature 
\cite{4,6,7,10,11}, in which some care must be taken because the
notation varies from  literature to literature.  Throughout this paper 
we use the notation as follows; a given one particle irreducible
$n$-point Feynman diagram in the single time representation contains a 
Keldish index at each end (external vertex) taking values \{1, 2\}, 
corresponding to the two branches of the closed-time-path contour,
while in the physical representation it contains a retarded-advanced 
index at each end taking values $\{R,\ A\}$, corresponding to the 
retarded or the advanced prescription. Our notation follows the one in 
Ref. \cite{7}. 

\subsection{The $n$-point vertex functions ($n$=2, 3, 4) in the
physical representation}

For convenience we here present explicitly the 2-, 3-, and 4-point
vertex functions in the physical representation constructed from the 
components in the single time representation.  Generally speaking the 
$n$-point function has $2^n$ components. These components obey one 
constraint equation, which reduces the number of independent
components to $2^n-1$. In the physical representation this is
expressed by the fact that the $n$-point vertex function with 
$n$-external advanced indices automatically vanishes,
Eqs. (\ref{3.1a}), (\ref{3.3a}) and (\ref{3.4a}), below. In 
equilibrium, the Kubo-Martin-Schwinger conditions impose additional 
constraints, reducing the number of independent components to 
$2^{(n-1)}-1$. For details of the construction, see Refs.\cite{4,7}.  
Since the transformation formula between the vertex functions in two 
representations do not take care of the external particle species 
(fermion or gauge boson), here we simply write 
$V_{\alpha \beta... \delta},\ \{\alpha,\beta,...,\delta = R,A\}$, or 
$V_{i j...l},\ \{i,j,...,l = 1,2\}$ for vertex functions in the 
physical and the single time representations, respectively. 
 
\subsubsection{2-point vertex functions}
{
\setcounter{enumi}{\value{equation}}
\addtocounter{enumi}{1}
\setcounter{equation}{0}
\renewcommand{\theequation}{\thesection.\theenumi\alph{equation}}
\begin{eqnarray}
\label{3.1a}   V_{AA} & = & 0,\\
\label{3.1b}
V_{RA} \ (\ \equiv V_R) & = &  V_{11}+V_{12},\\
\label{3.1c}
V_{AR} \ (\ \equiv V_A) & = & V_{11}+V_{21},\\
\label{3.1d}
V_{RR} \ (\ \equiv V_C) & = &  V_{11}+V_{22}.
\end{eqnarray}
\setcounter{equation}{\value{enumi}}}Among them the following relations hold,
\begin{subequations}
\begin{eqnarray}
\label{3.2a}
V_{AR}(K) & = & {V_{RA}}^*(K),\\
\label{3.2b}
V_{RR}(K) & = &  \{1+\eta 2n(k_0)\}\mbox{sgn}(k_0)\{V_{RA}(K) - V_{AR}(K)\},
\end{eqnarray}
\end{subequations}
where $\eta = +1(-1)$ for boson (fermion) and $n(k_0)$ is the 
corresponding equilibrium distribution function.

\subsubsection{3-point vertex functions}
\begin{subequations}
\begin{eqnarray}
\label{3.3a}V_{AAA} & = &  0,\\
\label{3.3b}
V_{RAA} & = &  V_{111}+V_{112}+V_{121}+V_{122},\\
\label{3.3c}
V_{ARA} & = &  V_{111}+V_{112}+V_{211}+V_{212},\\
\label{3.3d}
V_{AAR} & = &  V_{111}+V_{121}+V_{211}+V_{221},\\
\label{3.3e}
V_{RRA} & = &  V_{111}+V_{112}+V_{221}+V_{222},\\
\label{3.3f}
V_{RAR} & = &  V_{111}+V_{121}+V_{212}+V_{222},\\
\label{3.3g}
V_{ARR} & = &  V_{111}+V_{211}+V_{122}+V_{222},\\
\label{3.3h}
V_{RRR} & = &  V_{111}+V_{122}+V_{212}+V_{221}.
\end{eqnarray}
\end{subequations}
The Kubo-Martin-Schwinger conditions impose additional 4 
constraints, reducing the number of independent components to 3.

\subsubsection{4-point vertex functions}
\begin{subequations}
\begin{eqnarray}
\label{3.4a}V_{AAAA} & = &  0,\\
\label{3.4b}
V_{RAAA} & = & V_{1111}+V_{1112}+V_{1121}+V_{1211}
+V_{1122}+V_{1212}+V_{1221}+V_{1222},\\
\label{3.4c}
V_{ARAA} & = &  V_{1111}+V_{1112}+V_{1121}+V_{2111}
+V_{1122}+V_{2112}+V_{2121}+V_{2122},\\
\label{3.4d}
V_{AARA} & = &  V_{1111}+V_{1112}+V_{1211}+V_{2111}
+V_{1212}+V_{2112}+V_{2211}+V_{2212},\\
\label{3.4e}
V_{AAAR} & = &  V_{1111}+V_{1121}+V_{1211}+V_{2111}
+V_{1221}+V_{2121}+V_{2211}+V_{2221},\\
\label{3.4f}
V_{RRAA} & = &  V_{1111}+V_{1112}+V_{1121}+V_{1122}
+V_{2211}+V_{2212}+V_{2221}+V_{2222},\\
\label{3.4g}
V_{RARA} & = &  V_{1111}+V_{1112}+V_{1211}+V_{1212}
+V_{2121}+V_{2122}+V_{2221}+V_{2222},\\
\label{3.4h}
V_{RAAR} & = &  V_{1111}+V_{1121}+V_{1211}+V_{1221}
+V_{2112}+V_{2122}+V_{2212}+V_{2222}.
\end{eqnarray}
\end{subequations}
There are seven independent vertex functions, Eqs. (\ref{3.4a}-h), in 
this case, and the other eight vertex functions can be obtained from 
them using the Kubo-Martin-Schwinger conditions.

\subsection{ The $n$-point vertex functions ($n$=2, 3, 4) in the HTL 
approximation}
Now we calculate the $n$-point vertex functions ($n$=2, 3, 4) in the 
physical representation in the HTL approximation.  In the following we 
consider the massless QCD/QED, namely the high temperature hot
QCD/QED, where all the fermions and the gauge bosons are massless. 

Because we are accustomed to the Feynman rules in the single time
representation, we calculate explicitly the right-hand-side of
Eqs. (III.1), (III.3) and (III.4).  As we have stressed at the end of the last
section we use Eqs. (\ref{2.13}) and (II.14) for propagators, namely, we do not
use the expressions that contain explicitly the singular functions
themselves, but rather use the functions with the regularization
parameter $\varepsilon$ and keep $\varepsilon$ finite (i.e.,
$\varepsilon \neq 0$) till the end of all calculations then take the
limit  $\varepsilon \ \rightarrow\ 0$   at the end.   

\subsubsection{2-point vertex functions, or the fermion self-energy
and the gauge boson polarization tensor }

The 2-point vertex function in QCD/QED is usually named as the 
self-energy part $\Sigma$ for fermion, while it is called as the 
(vacuum) polarization tensor $\Pi^{\mu \nu}$ for gauge boson 
(gluon/photon). Although the HTL results of $\Sigma$ and 
$\Pi^{\mu \nu}$ in the single time representation as well as in the 
ITF are already well-known \cite{1,12,13}, for the sake of
completeness we here present the results in the physical 
representation. In fact in calculating their HTL contributions in the 
physical representation, some care being worth mentioning must be
taken, obtaining the result that could not be seen, at least in the 
explicit form, in the previous calculations \cite{1,12}.

\begin{figure}
\includegraphics{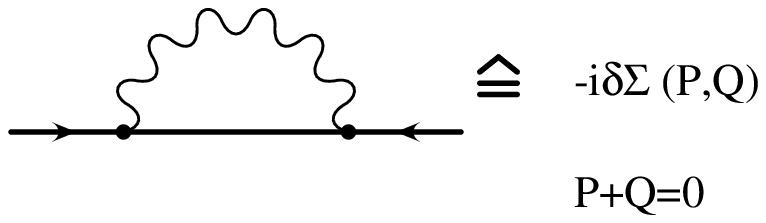}
\caption{\label{fig1}One-loop diagram for the fermion self-energy and
   the definition of its HTL, $\delta\Sigma$.
   $ \hat{=}$ denotes the equality in the HTL approximation.}
\end{figure}

The HTL contribution to the fermion self energy, $\delta \Sigma$, in 
the single time representation is obtained in QCD by calculating the 
diagram shown in Fig.\ref{fig1} (in QED $g^2 C_F$ should read $e^2$);
\begin{equation}
- i \delta \Sigma_{ij}(P,Q)    =
                       2g^2C_F\int\frac{d^4K}{(2\pi)^4} \not{\!K} \overline{D}_{ij}(K+Q)D_{ji}(K),\ \ \ P+Q=0,
\end{equation}   
where $D_{ij}$ and $\overline{D}_{ij}$ are given in Eqs.(II.2) and
(II.4).  With the use of Eqs.(III.1) we get the HTL fermion self
energy in the physical representation. There are nothing new in the 
obtained results in QCD, thus only reproducing the previous results;
\begin{subequations}
\begin{eqnarray}
\label{3.6a}
\delta \Sigma_{RA}(P,Q)  
&   = & - \frac{g^2C_F}{32\pi}T^2\int d\Omega \frac{\hat{\not{\!K}}}{
   Q\cdot\hat{K}+i \varepsilon},\\
\label{3.6b}
 \delta \Sigma_{AR}(P,Q) 
&   = &  - \frac{g^2C_F}{32\pi}T^2\int d\Omega \frac{\hat{\not{\!K}}}{
   Q\cdot\hat{K}- i \varepsilon},\\
\label{3.6c}
\delta \Sigma_{RR}(P,Q) & = &  0,
\end{eqnarray}
\end{subequations}
where $\hat{K}^{\mu} = (1,
\hat{\mbox{\boldmath{$k$}}})$ with
$\hat{\mbox{\boldmath{$k$}}}\ \equiv\ \mbox{\boldmath{$k$}}/k,\
k=\sqrt{{\mbox{\boldmath{$k$}}}^2}$, being the unit
three vector along the direction of $\mbox{\boldmath{$k$}}$. 

As was expressed in Eqs.(III.1) we usually denote the
$\{RA\}$-component as the $R$-, or the retarded component, and 
the $\{AR\}$-component as the $A$-, or the advanced component of 
the corresponding quantity. Namely $\Sigma_{RA} \equiv \Sigma_R$ 
is the fermion self-energy part of the inverse  retarded fermion 
propagator, having a definite physical meaning, i.e., the physical 
fermion mass function.

\begin{figure}
\includegraphics{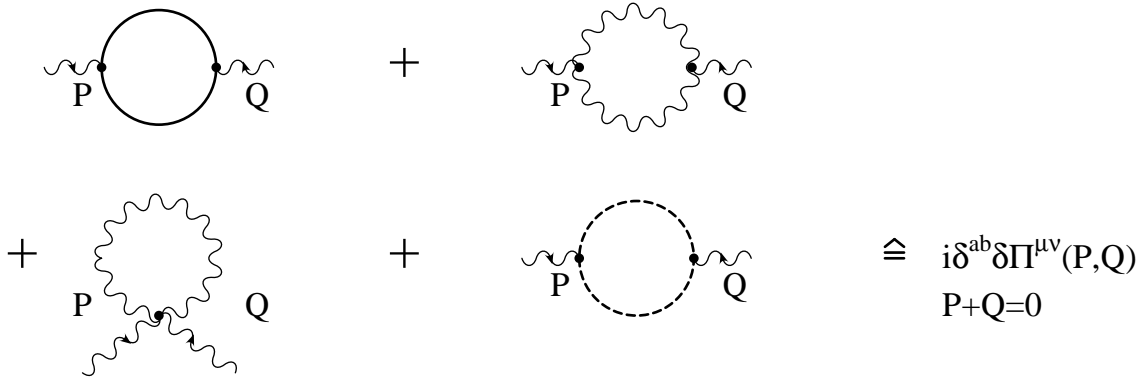}
\caption{\label{fig2}One-loop diagram for the gauge boson polarization
  tensor andthe definition of its HTL,
  $ \delta\Pi^{\mu \nu}$.  In QED only the diagram (a) exists.}
\end{figure}

As for the HTL contribution to the gauge boson polarization tensor, 
$\delta \Pi^{\mu \nu}$, the diagrams to be calculated in the single 
time representation are shown in Fig. \ref{fig2} $(P+Q=0)$ ;
\begin{equation}
i \delta \Pi^{\mu \nu}_{ij}(P,Q) =  \left\{
\begin{array}{ll}
-4 e^2 (-1)^{i+j-2}\int \frac{\displaystyle{d^4K}}{\displaystyle{(2\pi)^4}}
\left[2K^{\mu}K^{\nu}-g^{\mu\nu}K^2
\right]\overline{D}_{ij}(K)\overline{D}_{ji}(K+P) ,
 & \mbox{(QED)}\\
& \\
2g^2(-1)^{i+j-2} \int \frac{\displaystyle{d^4K}}{\displaystyle{(2\pi)^4}}
\left[2K^{\mu}K^{\nu}-g^{\mu\nu}K^2\right] & \\
\ \ \times \left\{N_c
{D}_{ij}(K){D}_{ji}(K+P) 
-N_f\overline{D}_{ij}(K)\overline{D}_{ji}(K+P)\right\}, &  (QCD)\\
\end{array}\right.
\end{equation}
from which we get the HTL gluon polarization tensor in QCD (for the 
photon polarization tensor in QED, $g^2(N_c+(1/2)N_f)$ should be 
understood as $e^2$),
\begin{subequations}
\begin{eqnarray}
\label{3.8a} 
 \delta \Pi^{\mu \nu}_{RA}(P,Q) & = & -\frac{g^2T^2}{12\pi}
\left(N_c+\frac{1}{2}N_f\right)
\int d\Omega \left({\hat{K}^{\mu}\hat{K}^{\nu}}\frac{q_0}{\hat{K}
{\cdot}Q + i \varepsilon}-g^{\mu 0}g^{\nu 0}\right),\\
\label{3.8b} 
\delta \Pi^{\mu \nu}_{AR}(P,Q) & = & -\frac{g^2T^2}{12\pi}
\left(N_c+\frac{1}{2}N_f\right)
\int d\Omega \left({\hat{K}^{\mu}\hat{K}^{\nu}}\frac{q_0}{\hat{K}
{\cdot}Q - i \varepsilon}-g^{\mu 0}g^{\nu 0}\right), \\
\label{3.8c} 
\delta \Pi^{\mu \nu}_{RR}(P,Q) & = & -\frac{g^2T^3}{6\pi}
\left(N_c+\frac{1}{2}N_f\right)
\int d\Omega \hat{K}^{\mu}\hat{K}^{\nu}
\left(\frac{1}{\hat{K}
{\cdot}Q + i \varepsilon}-\frac{1}{\hat{K}
{\cdot}Q - i \varepsilon}\right).
\end{eqnarray}
\end{subequations}
Same as the fermion self-energy function $\Pi^{\mu \nu}_{RA} 
(\Pi^{\mu \nu}_{AR}) \equiv \Pi^{\mu \nu}_R (\Pi^{\mu \nu}_A)$ is 
the gauge boson polarization tensor in the inverse retarded 
(advanced) gauge boson propagator, thus having a definite physical meaning.

It should be noted that the $\{RR\}$-component of the polarization 
tensor, $\delta \Pi^{\mu \nu}_{RR}$, is proportional to $g^2 T^3$, 
compared to the ordinary retarded or advanced polarization tensor, 
$\delta \Pi^{\mu \nu}_{RA}$ or $\delta \Pi^{\mu \nu}_{AR}$, being 
proportional to  $g^2 T^2$.  Also noted is that all the HTL
contribution in the ITF are of the order  $O(g^2 T^2)$, thus the 
above results show completely new vertex functions appear in the RTF.

\subsubsection{3-point vertex functions}

There are two types of 3-point vertex functions, the
fermion-gauge-boson vertex function and the 3-gauge-boson vertex 
function. Needless to say in QED only the fermion-gauge-boson vertex 
function exists. 

\vspace{0.5cm}
\paragraph{The fermion-gauge-boson vertex functions}

\begin{figure}
\includegraphics{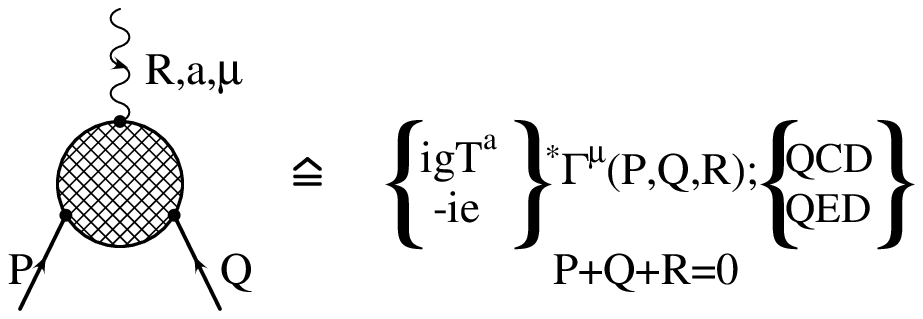}
\caption{\label{fig3}Definition of the fermion-gauge boson HTL
   resummed vertex function $\ ^*\Gamma^{\mu}$.}
\end{figure}

\begin{figure}
\includegraphics{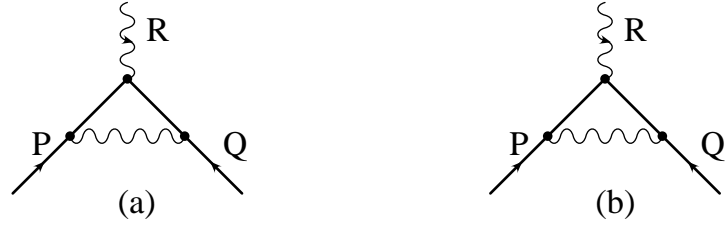}
\caption{\label{fig4}One-loop diagrams for the fermion-gauge boson
   vertex function.  In QED only the diagram (a) exists.}
\end{figure}

We define the fermion-gauge-boson vertex function in the HTL 
approximation as in Fig.\ref{fig3}, where
\begin{subequations}
\begin{equation}
\ ^{*}\Gamma^{\mu} =  \gamma^{\mu} + \delta \Gamma^{\mu}, 
\end{equation}
with $\gamma^{\mu}$ representing the tree vertex
\begin{eqnarray}
\gamma^{\mu}_{ijk} & = & (-)^{(i-1)}\gamma^{\mu},\ \ \  i=j=k,\nonumber \\
          &  = &   0 \ ,\ \ \ \ \ \ \ \ \ \ \ \ \ \ \ \      \mbox{otherwise},
\end{eqnarray}
\end{subequations}
and the HTL contribution to the
fermion-gauge-boson vertex function, $\delta \Gamma^{\mu}$, is obtained
by calculating the one-loop diagrams shown in Fig.\ref{fig4};
\begin{subequations}
\begin{eqnarray}
& & \delta \Gamma^{\mu}_{ijk}(P,Q,R) =   \left\{\begin{array}{ll}
4ie^2 \int \frac{\displaystyle{d^4K}}{\displaystyle{(2\pi)^4}}
K^\mu \not{\!K}\tilde{V}_{ijk}\ , &  \mbox{(QED)}\\
4ig^2C_F \int \frac{\displaystyle{d^4K}}{\displaystyle{(2\pi)^4}}
K^\mu \not{\!K}\tilde{V}_{ijk}
-2ig^2N_c \int \frac{\displaystyle{d^4K}}{\displaystyle{(2\pi)^4}}
K^\mu \not{\!K}\left(\tilde{V}_{ijk}+V_{ijk}\right)\ , &  \mbox{(QCD)}\\
\end{array}\right. 
\end{eqnarray}
where,
\begin{eqnarray}
& & V_{ijk} \ \equiv \ 
(-1)^{i+j+k-3}\overline{D}_{ij}(K)D_{jk}(K-Q)D_{ki}(K+P)\ ,\\
& & \tilde{V}_{ijk} \ \equiv \  (-1)^{i+j+k-3}{D}_{ij}(K)
\overline{D}_{jk}(K-Q)\overline{D}_{ki}(K+P)\ .
\end{eqnarray}
\end{subequations}
The results in QCD in the physical representation are given as follows 
(in QED $g^2 C_F$ should be understood as $e^2$);
\begin{subequations}
\begin{eqnarray}
\label{3.11a}
\delta \Gamma^{\mu}_{RAA}(P,Q,R)& = & -\frac{g^2T^2}{32\pi}C_F \int d\Omega
\frac{\hat{K}^\mu\not{\!\hat{K}}}{(\hat{K}{\cdot}P-i\varepsilon)
(\hat{K}{\cdot}Q+i\varepsilon)}\ ,\\
\label{3.11b}
\delta \Gamma^{\mu}_{ARA}(P,Q,R)& = & -\frac{g^2T^2}{32\pi}C_F \int d\Omega
\frac{\hat{K}^\mu\not{\!\hat{K}}}{(\hat{K}{\cdot}P+i\varepsilon)
(\hat{K}{\cdot}Q-i\varepsilon)}\ ,\\
\label{3.11c}
\delta \Gamma^{\mu}_{AAR}(P,Q,R) & = & -\frac{g^2T^2}{32\pi}C_F \int d\Omega
\frac{\hat{K}^\mu\not{\!\hat{K}}}{(\hat{K}{\cdot}P+i\varepsilon)
(\hat{K}{\cdot}Q+i\varepsilon)}\ ,\\
\label{3.11d}
\delta \Gamma^{\mu}_{RRR}(P,Q,R) & = &-\frac{g^2T^2}{32\pi}C_F \int d\Omega
\frac{\hat{K}^\mu\not{\!\hat{K}}}{(\hat{K}{\cdot}P-i\varepsilon)
(\hat{K}{\cdot}Q-i\varepsilon)}\ , \\
\label{3.11e}
\delta \Gamma^{\mu}_{RRA}(P,Q,R) & = &  \delta \Gamma^{\mu}_{RAR}(P,Q,R)
             = \delta \Gamma^{\mu}_{ARR}(P,Q,R)
= \delta \Gamma^{\mu}_{AAA}(P,Q,R)=0\ .
\end{eqnarray}
\end{subequations}
We can see that all the HTL terms of the fermion-gauge-boson vertex
function, $\delta \Gamma^{\mu}$, are proportional to $ g^2T^2$. 

\vspace{0.3cm}
In obtaining above results it is important to use the regularized
expressions for free propagators with the finite regularization
parameter $\varepsilon \ (\ \neq 0)$, Eqs. (\ref{2.1})-(II.4) and 
(\ref{2.13})-(II.14), not using the expressions in terms of the 
explicit singular functions from the beginning. To see this point 
more clearly let us calculate the above HTL vertex function 
$\delta\Gamma^{\mu}_{RAA}$,  Eq.(\ref{3.11a}), explicitly in QED. 
In each calculation of (\ref{3.11a}-d) we must evaluate four diagrams 
in the single time representation, Eqs.(III.3), thus face the 
calculation, e.g., of $\delta \Gamma^{\mu}_{111}(P,Q,R)$,
\begin{equation}
\label{3.12}
 \delta \Gamma^{\mu}_{111}(P,Q,R)=  4ie^2 \int \frac{d^4K}{(2\pi)^4}
{D}_{11}(K)\overline{D}_{11}(K-Q)\overline{D}_{11}(K+P).
\end{equation}
The loop-momentum integration over $K^{\mu}$ should be performed by 
keeping every $\varepsilon_i$ finite (i.e., $\varepsilon_i \neq 0$). 
Then the singularities of the integrand as a function of $k_0$ are
only poles in the complex $k_0$-plane thus the integration over $k_0$ 
can be carried out through the residue analysis. By neglecting the 
$O(e^2T)$ contributions we get after some manipulations (here we set 
$\varepsilon_1 = \varepsilon_2 = \varepsilon_3 = \varepsilon)$
\begin{equation}
\label{3.13}
\delta \Gamma^{\mu}_{111}(P,Q,R) =  -\frac{e^2}{(2\pi)^3 }
    \int dk k n(k) d\Omega K_{+}^{\mu}{\not{\!K}_{+}} 
                          \frac{K_{+}P}{(K_{+}P)^2 + (\varepsilon)^2}
      \frac{K_{+}Q}{(K_{+}Q)^2 + (\varepsilon)^2},
\end{equation}
where $n(k)=n_B(k)+n_F(k)$ and
$K_{+}^{\mu}=k(1,\hat{\mbox{\boldmath$k$}})
=k\hat{K}^{\mu}$ with $\hat{\mbox{\boldmath{$k$}}}= 
\mbox{\boldmath{$k$}}/k$.  Similar calculation gives
\begin{subequations}
\begin{eqnarray}
\label{3.14a}
\delta \Gamma^{\mu}_{112}(P,Q,R)& = &  -\frac{e^2}{(2\pi)^3 }
\int dk k n(k) d\Omega K_{+}^{\mu}{\not{\!K}_{+}} 
            \frac{\varepsilon}{(K_{+}P)^2 + (\varepsilon)^2}
            \frac{\varepsilon}{(K_{+}Q)^2 + (\varepsilon)^2}\ ,\\
\label{3.14b}
\delta \Gamma^{\mu}_{121}(P,Q,R) & = &  -i\frac{e^2}{(2\pi)^3 }
         \int dk k n(k) d\Omega K_{+}^{\mu}{\not{\!K}_{+}} 
                            \frac{K_{+}P}{(K_{+}P)^2 + (\varepsilon)^2}
                      \frac{\varepsilon}{(K_{+}Q)^2 + (\varepsilon)^2}\ ,\\
\label{3.14c}
\delta \Gamma^{\mu}_{122}(P,Q,R) & = &  i\frac{e^2}{(2\pi)^3 }
         \int dk k n(k) d\Omega K_{+}^{\mu}{\not{\!K}_{+}} 
            \frac{\varepsilon}{(K_{+}P)^2 + (\varepsilon)^2}
                       \frac{K_{+}Q}{(K_{+}Q)^2 + (\varepsilon)^2}\ .
\end{eqnarray}
\end{subequations}
Adding Eqs.(\ref{3.13}), (\ref{3.14a}-c) we get Eq.(\ref{3.11a}).

If we take naively the limit $\varepsilon_i \rightarrow 0$  before 
the loop-momentum integration over $K^{\mu}$, and use in 
Eq.(\ref{3.12}) the free propagators explicitly containing the 
singular Dirac-$\delta$ function and the principal part, then the 
naive manipulation gives, instead of Eq.(\ref{3.13}),
$$ \delta \Gamma^{\mu}_{111}(P,Q,R) =   -\frac{e^2}{(2\pi)^3}
                       \int dk k n(k) d\Omega K_{+}^{\mu}{\not{\!K}_{+}}  
     \left[ \mbox{PP}\frac1{K_{+}P}\mbox{PP}\frac1{K_{+}Q} +
          \frac{\pi^2}{3}\delta(K_{+}P)\delta(K_{+}Q)\right].
  \eqno{(\mbox{III.13'})} $$
Other three vertex functions coincide with the $\varepsilon_i
\rightarrow 0$ limit of Eqs.(\ref{3.14a}-c) and we can not get
Eq.(\ref{3.11a}). Additional term in Eq.(III.13') being proportional to
the product of Dirac-$\delta$ functions  may have its origin from the 
integral whose integrand being the product of two principal parts 
\begin{equation}
\int dy \int dx\mbox{PP}[\frac1{x-a(y)}] \mbox{PP}[\frac1{x-b(y)}].
\end{equation}
If, at some $y=y_0$ inside the integration range over $y$, $a(y)=b(y)$ 
happens then we face the integration over $x$ of the square of
principal part, which can not be well-defined. This is exactly what 
happens in the above calculation of Eq. (\ref{3.12}) with the free 
propagators explicitly containing the singular Dirac-$\delta$ function 
and the principal part.  Without concrete and consistent prescriptions 
how to treat the product of singular functions we can not get a
definite result, and the $\varepsilon$-regularization method gives
such a prescription.  As we have already noted throughout this paper 
we use the $\varepsilon$-regularized singular functions. 

\vspace{0.5cm}
\paragraph{Three gauge-boson (three gluon) vertex functions}

\begin{figure}
\includegraphics{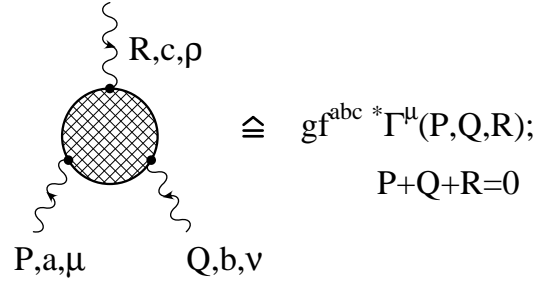}
\caption{\label{fig5}Definition of the 3-gauge boson HTL resummed
   vertex function $\ ^*\Gamma^{\mu \nu \rho}$.}
\end{figure}

\begin{figure}
\includegraphics{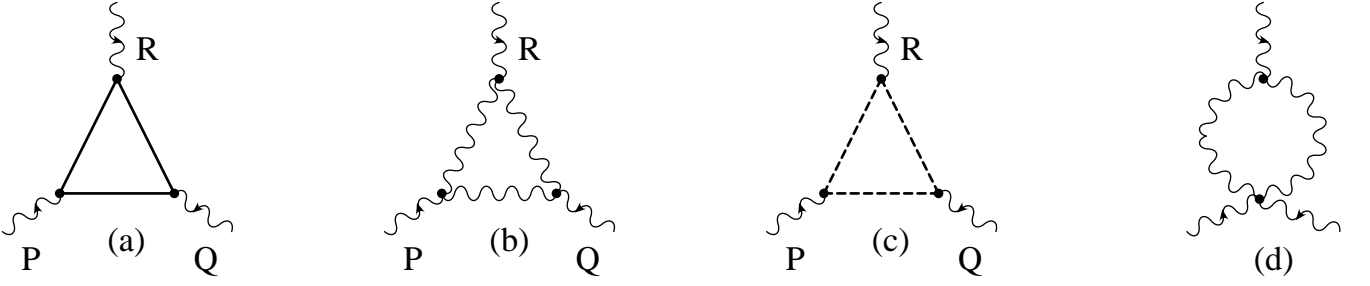}
\caption{\label{fig6}One-loop diagrams for the 3-gauge boson vertex
   function. Figure (d) represents all possible diagrams with independent
   configuration with respect to the three external legs with momenta 
   $P,\ Q$ and $R$.}
\end{figure}

We define the three gluon vertex function in the HTL approximation as 
in Fig.\ref{fig5}, where
\begin{equation}
\ ^*\Gamma^{\mu \nu \rho} = V^{\mu \nu \rho} + \delta \Gamma^{\mu \nu \rho},
\end{equation}
with  $V^{\mu \nu \rho}$ representing the tree vertex, which does not 
exist when the number of external retarded indices are zero or two.  
The HTL contribution to the three gluon vertex function, $\delta 
\Gamma^{\mu \nu \rho}$, is obtained by calculating the one-loop 
diagrams shown in Fig.\ref{fig6};
\begin{eqnarray}
 \delta \Gamma^{\mu \nu \rho}_{ijk}(P,Q,R) & = & 
-\frac{ig^2}{2\pi^4}(-1)^{i+k+j-3}
\int d^4K K^\mu K^\nu K^\rho \nonumber \\
& \times & 
\left[
N_cD_{ij}(K)D_{jk}(K-Q)D_{ki}(K+P)
-N_f\overline{D}_{ij}(K)\overline{D}_{jk}(K-Q)\overline{D}_{ki}(K+P)\right].
\end{eqnarray}
By using Eqs.(III.3) we get the HTL results in the physical representation;
\begin{subequations}
\begin{eqnarray}
\delta \Gamma^{\mu \nu \rho}_{RAA}(P,Q,R) & = & 
-\frac{g^2T^2}{12\pi}\left(N_c+\frac{1}{2}N_f\right) \int 
d\Omega \hat{K}^\mu \hat{K}^\nu \hat{K}^\rho 
\left[
\frac{p_0}{(\hat{K}{\cdot}P-i\varepsilon)(\hat{K}{\cdot}R+i\varepsilon)}
-\frac{q_0}{(\hat{K}{\cdot}Q+i\varepsilon)(\hat{K}{\cdot}R+i\varepsilon)}
\right], \nonumber \\
\label{3.18a}& & \\
\delta \Gamma^{\mu \nu \rho}_{ARA}(P,Q,R) & = & 
-\frac{g^2T^2}{12\pi}\left(N_c+\frac{1}{2}N_f\right) \int 
d\Omega \hat{K}^\mu \hat{K}^\nu \hat{K}^\rho 
\left[
\frac{p_0}{(\hat{K}{\cdot}P+i\varepsilon)(\hat{K}{\cdot}R+i\varepsilon)}
-\frac{q_0}{(\hat{K}{\cdot}Q-i\varepsilon)(\hat{K}{\cdot}R+i\varepsilon)}
\right], \nonumber \\
\label{3.18b} & & \\
\delta \Gamma^{\mu \nu \rho}_{AAR}(P,Q,R) & = & 
-\frac{g^2T^2}{12\pi}\left(N_c+\frac{1}{2}N_f\right) \int 
d\Omega \hat{K}^\mu \hat{K}^\nu \hat{K}^\rho 
\left[
\frac{p_0}{(\hat{K}{\cdot}P+i\varepsilon)(\hat{K}{\cdot}R-i\varepsilon)}
-\frac{q_0}{(\hat{K}{\cdot}Q+i\varepsilon)(\hat{K}{\cdot}R-i\varepsilon)}
\right],\nonumber  \\
\label{3.18c}& & \\
\delta \Gamma^{\mu \nu \rho}_{RRR}(P,Q,R) & = & 
-\frac{g^2T^2}{12\pi}\left(N_c+\frac{1}{2}N_f\right) \int 
d\Omega \hat{K}^\mu \hat{K}^\nu \hat{K}^\rho 
\left[
\frac{p_0}{(\hat{K}{\cdot}P-i\varepsilon)(\hat{K}{\cdot}R-i\varepsilon)}
-\frac{q_0}{(\hat{K}{\cdot}Q-i\varepsilon)(\hat{K}{\cdot}R-i\varepsilon)}
\right],\nonumber  \\
\label{3.18d} & &  \\
\delta \Gamma^{\mu \nu \rho}_{RRA}(P,Q,R) & = & 
-\frac{g^2T^3}{6\pi}\left(N_c+\frac{1}{2}N_f\right) \int 
d\Omega \hat{K}^\mu \hat{K}^\nu \hat{K}^\rho 
\left[
\frac{1}{(\hat{K}{\cdot}P-i\varepsilon)(\hat{K}{\cdot}Q+i\varepsilon)}
-\frac{1}{(\hat{K}{\cdot}P+i\varepsilon)(\hat{K}{\cdot}Q-i\varepsilon)}
\right],\nonumber   \\
\label{3.18e} & & \\
\delta \Gamma^{\mu \nu \rho}_{RAR}(P,Q,R) & = & 
-\frac{g^2T^3}{6\pi}\left(N_c+\frac{1}{2}N_f\right) \int 
d\Omega \hat{K}^\mu \hat{K}^\nu \hat{K}^\rho 
\left[
\frac{1}{(\hat{K}{\cdot}P+i\varepsilon)(\hat{K}{\cdot}R-i\varepsilon)}
-\frac{1}{(\hat{K}{\cdot}P-i\varepsilon)(\hat{K}{\cdot}R+i\varepsilon)}
\right],\nonumber   \\
\label{3.18f} & & \\
\delta \Gamma^{\mu \nu \rho}_{ARR}(P,Q,R) & = & 
-\frac{g^2T^3}{6\pi}\left(N_c+\frac{1}{2}N_f\right) \int 
d\Omega \hat{K}^\mu \hat{K}^\nu \hat{K}^\rho 
\left[
\frac{1}{(\hat{K}{\cdot}Q-i\varepsilon)(\hat{K}{\cdot}R+i\varepsilon)}
-\frac{1}{(\hat{K}{\cdot}Q+i\varepsilon)(\hat{K}{\cdot}R-i\varepsilon)}
\right],\nonumber  \\
\label{3.18g} & & \\
\label{3.18h}
\delta \Gamma^{\mu \nu \rho}_{AAA}(P,Q,R) & = & 0.
\end{eqnarray}
\end{subequations}
It should be noted that depending on the number of external retarded 
indices the HTL three gluon vertex functions can be classified into 
two groups; i) number of $R$  is 1 or 3, and ii)  number of $R$ is 0 
or 2. Any vertex belonging to the first group has the tree vertex and 
is proportional to $g^2 T^2$, while other vertices belonging to the 
second group do not have the tree terms and are proportional to
$g^2T^3$, having no counterparts in the ITF. 

\subsubsection{4-point vertex functions}
There are two types of 4-point vertex functions, the fermion pair and 
2-gauge boson vertex functions and the 4-gauge boson vertex
functions. In QED, as we shall see below, all the HTL contributions 
to the 4-photon vertex functions vanish. 

\vspace{0.5cm}
\paragraph{One fermion pair and 2-gauge boson vertex functions}

\begin{figure}
\includegraphics{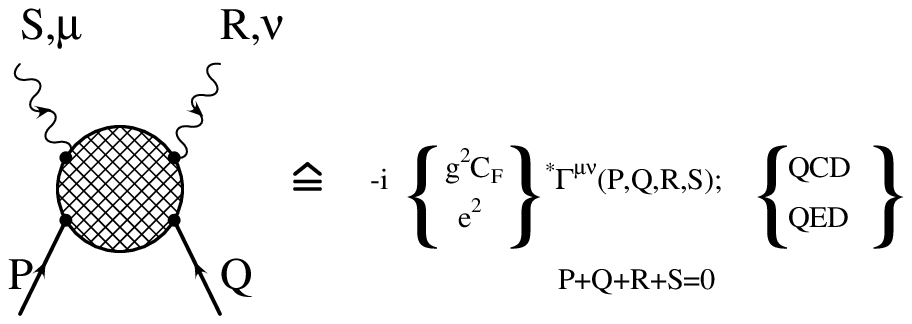}
\caption{\label{fig7}Definition of the fermion-pair-2-gauge boson HTL
    resummed vertex function $\ ^*\Gamma^{\mu  \nu}.$}
\end{figure}

\begin{figure}
\includegraphics{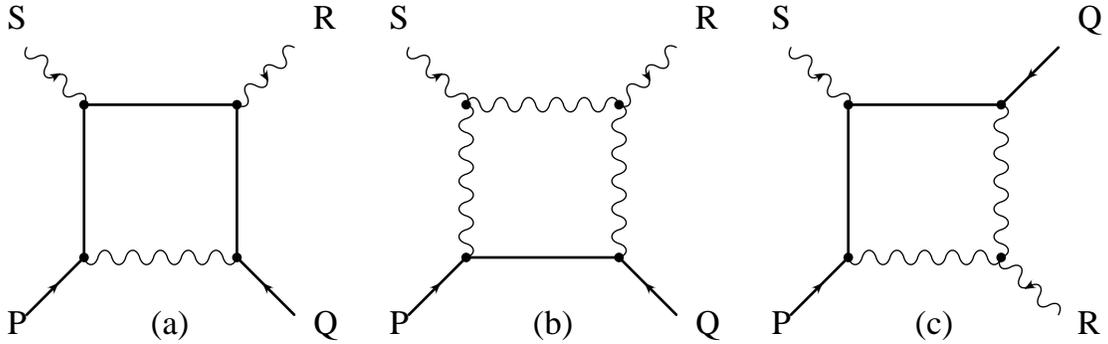}
\caption{\label{fig8}One-loop diagrams for the fermion-pair-2-gauge
   boson vertex function. Their possible exchanged diagrams between
   the external gauge boson legs with momenta $R$ and $S$ should also
   be added. In QED only the diagram (a) and its exchanged diagram exist.}
\end{figure}

We define the fermion pair-2-gauge boson vertex functions in the HTL 
approximation as in Fig.\ref{fig7}, where
\begin{equation}
\ ^*\Gamma^{\mu \nu} = \delta \Gamma^{\mu \nu},
\end{equation}
and the HTL contribution to the fermion pair-2-gauge boson vertex
function, $\delta \Gamma^{\mu \nu}$, is obtained by calculating the 
one-loop diagrams shown in Fig.\ref{fig8} and their exchanged 
diagrams between the external legs with momenta $R$ and $S$;
\begin{equation}
 \delta \Gamma^{\mu \nu}_{ijkl}(P,Q,R,S) = \left\{\begin{array}{l}
-i8e^2 (-1)^{i+j+k+l}\int \frac{\displaystyle{d^4K}}{\displaystyle{(2\pi)^4}}
K^\mu K^\nu \not{\!K}
{D}_{ij}(K)\overline{D}_{jk}(K-Q)\overline{D}_{kl}(K+P+S)
\overline{D}_{li}(K+P),\\
\hspace{11.9cm}  \mbox{(QED)}\\
-i8g^2C_F (-1)^{i+j+k+l}
\int \frac{\displaystyle{d^4K}}{\displaystyle{(2\pi)^4}} 
K^\mu K^\nu \not{\!K} \\
\hspace{3cm} \times
\left[C_F{D}_{ij}(K)\overline{D}_{jk}(K-Q)\overline{D}_{kl}(K+P+S)
\overline{D}_{li}(K+P)\right.  \\
\hspace{3cm} 
-N_c \overline{D}_{ij}(K)D_{jk}(K-Q)D_{kl}(K+P+S)D_{li}(K+P) \\
\left.\hspace{3cm} 
+\frac{\displaystyle{1}}{\displaystyle{2}}N_c{D}_{ij}(K+P)D_{jl}(K-P-S)
\overline{D}_{lk}(K+R)\overline{D}_{ki}(K)\right]. \ \mbox{(QCD)}\\
\end{array}\right.
\end{equation}
The results in QCD in the physical representation are, with the use of 
Eqs.(III.4), given as follows (in QED $g^2 C_F$ should be understood 
as $e^2$);
\begin{subequations}
\begin{eqnarray}
\label{3.21a}
\delta \Gamma^{\mu \nu}_{RAAA}(P,Q,R,S) & = & 
\frac{g^2T^2}{32\pi}C_F \int d\Omega \hat{K}^\mu\hat{K}^\nu\not{\!\hat{K}}
\frac{1}{\hat{K}{\cdot}P-i\varepsilon}\frac{1}{\hat{K}{\cdot}Q+i\varepsilon}
\left[\frac{1}{\hat{K}{\cdot}(P+S)-i\varepsilon}
+\frac{1}{\hat{K}{\cdot}(P+R)-i\varepsilon}\right], \nonumber \\
& & \\
\label{3.21b}\delta \Gamma^{\mu \nu}_{ARAA}(P,Q,R,S) & = & 
\frac{g^2T^2}{32\pi}C_F \int d\Omega \hat{K}^\mu\hat{K}^\nu\not{\!\hat{K}}
\frac{1}{\hat{K}{\cdot}P+i\varepsilon}\frac{1}{\hat{K}{\cdot}Q-i\varepsilon}
\left[\frac{1}{\hat{K}{\cdot}(P+S)+i\varepsilon}
+\frac{1}{\hat{K}{\cdot}(P+R)+i\varepsilon}\right], \nonumber \\
& & \\
\label{3.21c}\delta \Gamma^{\mu \nu}_{AARA}(P,Q,R,S) & = & 
\frac{g^2T^2}{32\pi}C_F \int d\Omega \hat{K}^\mu\hat{K}^\nu\not{\!\hat{K}}
\frac{1}{\hat{K}{\cdot}P+i\varepsilon}\frac{1}{\hat{K}{\cdot}Q+i\varepsilon}
\left[\frac{1}{\hat{K}{\cdot}(P+S)+i\varepsilon}
+\frac{1}{\hat{K}{\cdot}(P+R)-i\varepsilon}\right], \nonumber \\
& & \\
\label{3.21d}\delta \Gamma^{\mu \nu}_{AAAR}(P,Q,R,S)& = & 
\frac{g^2T^2}{32\pi}C_F \int d\Omega \hat{K}^\mu\hat{K}^\nu\not{\!\hat{K}}
\frac{1}{\hat{K}{\cdot}P+i\varepsilon}\frac{1}{\hat{K}{\cdot}Q+i\varepsilon}
\left[\frac{1}{\hat{K}{\cdot}(P+S)-i\varepsilon}
+\frac{1}{\hat{K}{\cdot}(P+R)+i\varepsilon}\right], \nonumber \\
& & \\
\label{3.21e}
\delta \Gamma^{\mu \nu}_{RRAA}(P,Q,R,S)& = & 
                  \delta \Gamma^{\mu \nu}_{RARA}(P,Q,R,S)
=\delta \Gamma^{\mu \nu}_{RAAR}(P,Q,R,S)=0.
\end{eqnarray}
\end{subequations}
Other eight vertex functions can be obtained from Eqs. (\ref{3.21a}-e) 
using the KMS conditions.  Again all the HTL contributions to the 
vertex functions with external fermion legs are of $O(g^2T^2)$.
 
Some comment on Eq.(\ref{3.21e}) must be given. Hou Defu et al, 
Ref.\cite{11}, calculated the same HTL vertex functions and claimed 
the existence of the non-zero HTL contributions, only the first 
function $\delta \Gamma^{\mu \nu}_{RRAA}(P,Q,R,S)$ to vanish in 
equilibrium. However, as we explicitly show, all the three functions 
vanish in the HTL approximation so long as the propagation of thermal 
"quasi-particles" can be described by the form of free particle 
propagators, Eqs. (\ref{2.1})-(II.4), no matter which in equilibrium 
nor just out-of-equilibrium. 

\vspace{0.5cm}
\paragraph{4-gauge boson vertex functions}

The HTL contribution to the 4-photon vertex function in QED completely 
vanishes, thus we confine our interest to the 4-gluon vertex functions 
in QCD.  Because of the complexity in their tensor structure and of
our interest in their application, we calculate the HTL's for the 
vertex functions being summed in the color indices over two of the 
external gluon legs. 

\begin{figure}
\includegraphics{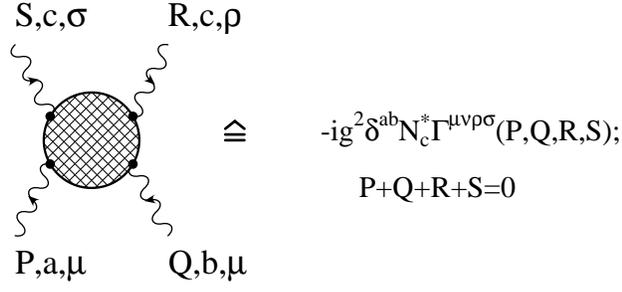}
\caption{\label{fig9}Definition of the 4-gluon HTL resummed vertex
   function $\ ^*\Gamma^{\mu\nu \rho \sigma}.$
   Color indices of the external gluons with momenta $R$ and $S$ is
   summed over, which is indicated by the same color index  $c$.}
\end{figure}

\begin{figure}
\includegraphics{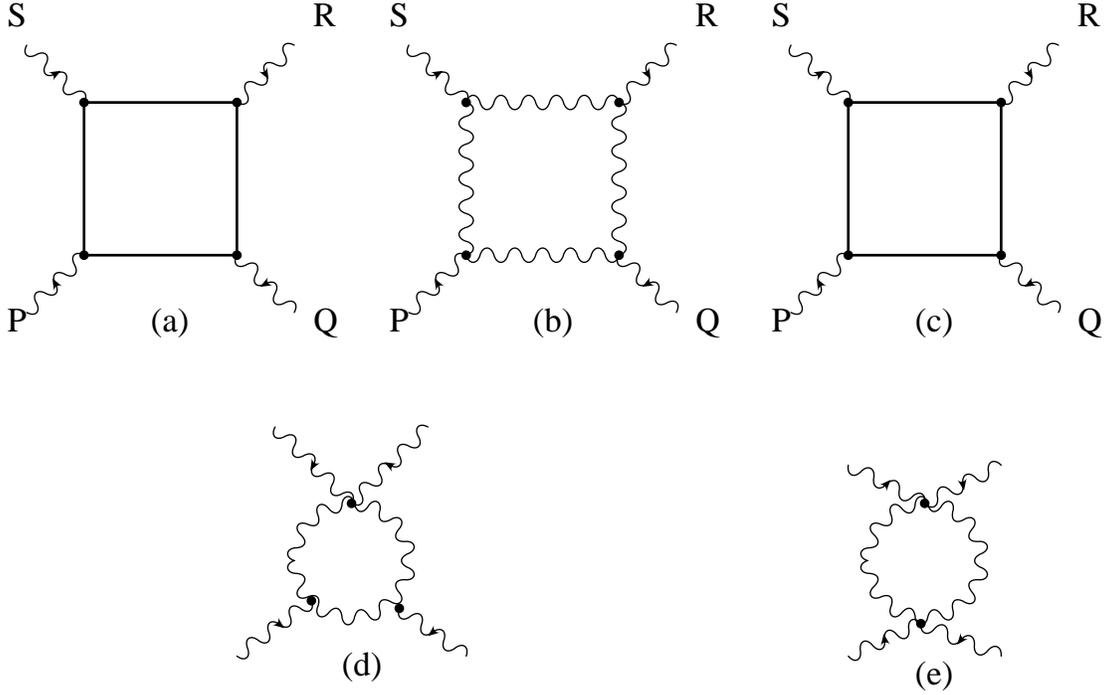}
\caption{\label{fig10}One-loop diagrams for the 4-gluon vertex
  function. For figures (a), (b) and (c), their possible exchanged
  diagrams among the external legs with momenta $Q,\ R$ and $S$ should
  also be added. Figures (d) and (e) represent all possible diagrams
  with independent configuration with respect to the four external
  legs with momenta $P, \ Q,\ R$ and $S$.}
\end{figure}

Defining the 4-gluon vertex functions in the HTL approximation as in 
Fig.\ref{fig9}, where the summation over the color indices of the 
gluons having momenta $R$ and $S$ should be understood and 
\begin{subequations}
\begin{equation}
\ ^*\Gamma^{\mu \nu \rho \sigma} =  
         W^{\mu \nu \rho \sigma}+\delta \Gamma^{\mu \nu \rho \sigma},
\end{equation}
with $ W^{\mu \nu \rho \sigma}$  representing the tree vertex,
\begin{eqnarray}
 W^{\mu \nu \rho \sigma}_{ijkl} & = & (-)^{i-1}
(2g^{\mu \nu}g^{\rho \sigma}-g^{\mu\rho}g^{\nu\sigma}
-g^{\mu\sigma}g^{\nu\rho}),\ \    i=j=k=l,\nonumber \\
&  = & 0,\hspace{5.5cm}      \mbox{otherwise}.
\end{eqnarray}
\end{subequations}
The HTL contribution to the 4-gluon
vertex function, $\delta \Gamma^{\mu \nu \rho \sigma}$, is obtained by
calculating the one-loop diagrams shown in Fig.\ref{fig10} and their all
possible exchanged diagrams among the external legs with momenta $Q,\ R$
and $S$ (Note that the color indices of the gluons with momenta $R$ and
$S$ are summed over);
\begin{subequations}
\begin{eqnarray}
& & \delta \Gamma^{\mu \nu \rho \sigma}_{ijkl}(P,Q,R,S)   =  
-g^2\frac{32}{N_c} \int \frac{d^4K}{(2\pi)^4}K^\mu K^\nu K^\rho K^\sigma
\nonumber \\
& & \ \times  
\left[N_c^2\left\{V_{ijkl}(P,Q,R,S)+V_{ijlk}(P,Q,S,R)\right\} 
\right.
-
C_FN_F\left\{\tilde{V}_{ijkl}(P,Q,R,S) 
+\tilde{V}_{ijlk}(P,Q,S,R)\right\}\nonumber \\
& & \left.\ \ \ 
+\frac{N_c^2}{2}V_{ikjl}(P,R,Q,S)+\frac{N_f}{2N_c}\tilde{V}_{ikjl}(P,R,Q,S)
\right],
\end{eqnarray}
where,
\begin{eqnarray}
& & V_{ijkl} \  \equiv \ 
 (-1)^{i+j+k+l}D_{ij}(K)D_{jk}(K-Q)D_{kl}(K+P+S)D_{li}(K+P),
\\
& & \tilde{V}_{ijkl} \ \equiv \ (-1)^{i+j+k+l}\overline{D}_{ij}(K)
\overline{D}_{jk}(K-Q)\overline{D}_{kl}(K+P+S)\overline{D}_{li}(K+P).
\end{eqnarray}
\end{subequations}
The results in QCD in the physical representation are, with the use 
of Eqs.(III.4), given as follows;
\begin{subequations}
\begin{eqnarray}
\delta \Gamma^{\mu \nu \rho \sigma}_{RAAA}(P,Q,R,S) & = & 
-\frac{g^2T^2}{6\pi}\left(N_c+\frac{1}{2}N_f\right)
\int d\Omega \hat{K}^\mu\hat{K}^\nu\hat{K}^\rho\hat{K}^\sigma 
\frac{1}{\hat{K}{\cdot}S+i\varepsilon}
\frac{1}{\hat{K}{\cdot}R+i\varepsilon}\nonumber \\
\label{3.24a}& \times & 
\left[\frac{p_0+r_0}{\hat{K}{\cdot}(P+R)-i\varepsilon}
-\frac{q_0}{\hat{K}{\cdot}Q-i\varepsilon}
-\frac{p_0}{\hat{K}{\cdot}P-i\varepsilon}
+\frac{q_0+r_0}{\hat{K}{\cdot}(Q+R)+i\varepsilon}
\right],\\
\delta \Gamma^{\mu \nu \rho \sigma}_{ARAA}(P,Q,R,S)  & = & 
-\frac{g^2T^2}{6\pi}\left(N_c+\frac{1}{2}N_f\right)
\int d\Omega \hat{K}^\mu\hat{K}^\nu\hat{K}^\rho\hat{K}^\sigma 
\frac{1}{\hat{K}{\cdot}S+i\varepsilon}
\frac{1}{\hat{K}{\cdot}R+i\varepsilon}\nonumber \\
\label{3.24b}& \times & 
\left[\frac{p_0+r_0}{\hat{K}{\cdot}(P+R)+i\varepsilon}
-\frac{q_0}{\hat{K}{\cdot}Q-i\varepsilon}
-\frac{p_0}{\hat{K}{\cdot}P+i\varepsilon}
+\frac{q_0+r_0}{\hat{K}{\cdot}(Q+R)-i\varepsilon}
\right],\\
\delta \Gamma^{\mu \nu \rho \sigma}_{AARA}(P,Q,R,S)  & = & 
-\frac{g^2T^2}{6\pi}\left(N_c+\frac{1}{2}N_f\right)
\int d\Omega \hat{K}^\mu\hat{K}^\nu\hat{K}^\rho\hat{K}^\sigma 
\frac{1}{\hat{K}{\cdot}S+i\varepsilon}
\frac{1}{\hat{K}{\cdot}R-i\varepsilon}\nonumber \\
\label{3.24c}& \times & 
\left[\frac{p_0+r_0}{\hat{K}{\cdot}(P+R)-i\varepsilon}
-\frac{q_0}{\hat{K}{\cdot}Q+i\varepsilon}
-\frac{p_0}{\hat{K}{\cdot}P+i\varepsilon}
+\frac{q_0+r_0}{\hat{K}{\cdot}(Q+R)-i\varepsilon}
\right],\\
\delta \Gamma^{\mu \nu \rho \sigma}_{AAAR}(P,Q,R,S) & = & 
-\frac{g^2T^2}{6\pi}\left(N_c+\frac{1}{2}N_f\right)
\int d\Omega \hat{K}^\mu\hat{K}^\nu\hat{K}^\rho\hat{K}^\sigma 
\frac{1}{\hat{K}{\cdot}S-i\varepsilon}
\frac{1}{\hat{K}{\cdot}R+i\varepsilon}\nonumber \\
\label{3.24d}& \times & 
\left[\frac{p_0+r_0}{\hat{K}{\cdot}(P+R)+i\varepsilon}
-\frac{q_0}{\hat{K}{\cdot}Q+i\varepsilon}
-\frac{p_0}{\hat{K}{\cdot}P+i\varepsilon}
+\frac{q_0+r_0}{\hat{K}{\cdot}(Q+R)+i\varepsilon}
\right],\\
\delta \Gamma^{\mu \nu \rho \sigma}_{RRAA}(P,Q,R,S) & = & 
\frac{g^2T^3}{6\pi}\left(N_c+\frac{1}{2}N_f\right)
\int d\Omega \hat{K}^\mu\hat{K}^\nu\hat{K}^\rho\hat{K}^\sigma
(P-Q)\cdot\hat{K} \nonumber \\
\label{3.24e}& \times & 
\left[\frac{1}{\hat{K}{\cdot}(P+R)+i\varepsilon}
\frac{1}{\hat{K}{\cdot}P+i\varepsilon}
\frac{1}{\hat{K}{\cdot}(Q+R)-i\varepsilon}
\frac{1}{\hat{K}{\cdot}Q-i\varepsilon}-c.c
\right],\\
\delta \Gamma^{\mu \nu \rho \sigma}_{RARA}(P,Q,R,S) & = & 
-\frac{g^2T^3}{6\pi}\left(N_c+\frac{1}{2}N_f\right)
\int d\Omega \hat{K}^\mu\hat{K}^\nu\hat{K}^\rho\hat{K}^\sigma \nonumber \\
\label{3.24f} & \times & \left[\frac{1}{\hat{K}{\cdot}P+i\varepsilon}
\frac{1}{\hat{K}{\cdot}(Q+R)-i\varepsilon}
\frac{1}{\hat{K}{\cdot}R-i\varepsilon}-c.c
\right],\\
\delta \Gamma^{\mu \nu \rho \sigma}_{RAAR}(P,Q,R,S) & = & 
-\frac{g^2T^3}{6\pi}\left(N_c+\frac{1}{2}N_f\right)
\int d\Omega \hat{K}^\mu\hat{K}^\nu\hat{K}^\rho\hat{K}^\sigma \nonumber \\
\label{3.24g} & \times & \left[\frac{1}{\hat{K}{\cdot}P-i\varepsilon}
\frac{1}{\hat{K}{\cdot}(P+R)-i\varepsilon}
\frac{1}{\hat{K}{\cdot}S+i\varepsilon}-c.c
\right].
\end{eqnarray}
\end{subequations}
Other eight vertex functions can be obtained from Eqs. (\ref{3.24a}-g) 
using the KMS conditions.  It is worth noting that in this case again 
depending on the number of external retarded indices the HTL four
gluon vertex functions can be classified into two groups; i) number of 
$R$ is 1 or 3, and ii)  number of $R$ is 0 or 2. Any vertex belonging 
to the first group is of $O(g^2 T^2)$, while other vertices belonging 
to the second group are are of $O(g^2T^3)$, having no counterparts in 
the ITF. The existence of $O(g^2T^3)$ HTL contribution is
characteristic of the $n$-gauge boson vertex functions.

\section{Ward-Takahashi Identities in the HTL Approximation}
\label{s4}

Having determined the 2-, 3-, and 4-point vertex functions in QCD and 
in QED in the HTL approximation, in this section we verify that the 
Ward-Takahashi identities satisfied between them are all of the simple 
QED-type identities. As we have shown the HTL $n$-gauge boson vertex 
functions with even number external retarded indices show, contrasting
to the ordinary $O(g^2T^2$) behavior of other vertex functions, the 
$O(g^2T^3)$ behavior being totally absent in any amplitude in the
ITF. We here verify that they also satisfy the simple QED-type 
Ward-Takahashi identities between themselves in the HTL approximation, 
thus verifying with an explicit calculation the gauge invariance of 
thermal QCD/QED in the HTL approximation in the RTF, being guaranteed 
by the QED-type Ward-Takahashi identities.

\subsection{Ward-Takahashi identities between the three- and
four-point vertex functions in QCD/QED}

\subsubsection{Between the 3- and 4-point HTL vertex functions with 
a fermion pair legs}
There are no differences between the 3- and 4-point HTL vertex
functions with a fermion pair legs in QCD and in QED except the group 
factor, thus we only study in QCD.  Comparing the results Eqs.(III.21) 
and Eqs.(III.11) we obtain the following Ward-Takahashi identities 
between vertices shown in Fig.\ref{fig7} (i.e., \ref{fig8}) and 
Fig.\ref{fig3} (i.e., \ref{fig4}):
\begin{subequations}
\begin{eqnarray}
\label{4.1a}
R_{\mu}\ ^*\Gamma^{\mu \nu}_{RAAA}(P,Q,R,S) & = & 
\delta \Gamma^{\nu}_{RAA}(P+R,Q,S) - \delta \Gamma^{\nu}_{RAA}(P,Q+R,S),\\
\label{4.1b}
R_{\mu}\ ^*\Gamma^{\mu \nu}_{ARAA}(P,Q,R,S)& = & 
 \delta \Gamma^{\nu}_{ARA}(P+R,Q,S) - \delta \Gamma^{\nu}_{ARA}(P,Q+R,S),\\
\label{4.1c}
R_{\mu}\ ^*\Gamma^{\mu \nu}_{AARA}(P,Q,R,S) & = & 
   \delta \Gamma^{\nu}_{RAA}(P+R,Q,S) - \delta \Gamma^{\nu}_{ARA}(P,Q+R,S),\\
\label{4.1d}
R_{\mu}\ ^*\Gamma^{\mu \nu}_{AAAR}(P,Q,R,S) & = & 
  \delta \Gamma^{\nu}_{AAR}(P+R,Q,S) - \delta \Gamma^{\nu}_{AAR}(P,Q+R,S).
\end{eqnarray}
\end{subequations}
When we operate $S_{\mu}$ in place of $R_{\mu}$, then obviously the
role of $S$ and $R$ are exchanged in Eqs.(IV.1) and also the role of 
Eqs.(\ref{4.1c}) and (\ref{4.1d}) should be exchanged. 

\subsubsection{Between the 3- and 4-gluon HTL vertex functions}
In QED there are no 3-photon vertex functions and the 4-photon vertex 
functions totally vanish in  the HTL approximation, thus we study in 
QCD the relation between the 3- and 4-gluon HTL vertex functions. 
Comparing the results Eqs.(III.24) and Eqs.(III.18) we obtain the 
following Ward-Takahashi identities between vertices shown in 
Fig.\ref{fig9} (i.e., \ref{fig10}) and Fig.\ref{fig5} (i.e., \ref{fig6}):
\begin{subequations}
\begin{eqnarray}
\label{4.2a}
R_{\mu}\delta \Gamma^{\mu \nu \rho \sigma}_{RAAA}(P,Q,R,S)&
                    =& \delta \Gamma^{\nu \rho \sigma}_{RAA}(P+R,Q,S)
    - \delta \Gamma^{\nu \rho \sigma}_{RAA}(P,Q+R,S),\\
\label{4.2b}
R_{\mu}\delta \Gamma^{\mu \nu \rho \sigma}_{ARAA}(P,Q,R,S)&
                    = & \delta \Gamma^{\nu \rho \sigma}_{ARA}(P+R,Q,S)
       - \delta \Gamma^{\nu \rho \sigma}_{ARA}(P,Q+R,S), \\
\label{4.2c}
R_{\mu}\delta \Gamma^{\mu \nu \rho \sigma}_{AARA}(P,Q,R,S) &
                    = & \delta \Gamma^{\nu \rho \sigma}_{RAA}(P+R,Q,S)
      - \delta \Gamma^{\nu \rho \sigma}_{ARA}(P,Q+R,S),\\
\label{4.2d}
R_{\mu}\delta \Gamma^{\mu \nu \rho \sigma}_{AAAR}(P,Q,R,S) &
                    = & \delta \Gamma^{\nu \rho \sigma}_{AAR}(P+R,Q,S)
       - \delta \Gamma^{\nu \rho \sigma}_{AAR}(P,Q+R,S), \\
\label{4.2e}
R_{\mu}\delta \Gamma^{\mu \nu \rho \sigma}_{RRAA}(P,Q,R,S) & 
                    = & \delta \Gamma^{\nu \rho \sigma}_{RRA}(P+R,Q,S)
  - \delta \Gamma^{\nu \rho \sigma}_{RRA}(P,Q+R,S), \\
\label{4.2f}
R_{\mu}\delta \Gamma^{\mu \nu \rho \sigma}_{RARA}(P,Q,R,S) & 
                    = &  - \delta \Gamma^{\nu \rho \sigma}_{RRA}(P,Q+R,S), \\
\label{4.2g}
R_{\mu}\delta \Gamma^{\mu \nu \rho \sigma}_{RAAR}(P,Q,R,S) & 
                    = & \delta \Gamma^{\nu \rho \sigma}_{RAR}(P+R,Q,S)
   - \delta \Gamma^{\nu \rho \sigma}_{RAR}(P,Q+R,S).
\end{eqnarray}
\end{subequations}
When we operate $S_{\mu}$ in place of $R_{\mu}$, then obviously the 
role of $S$ and $R$  are exchanged in Eqs.(IV.2) and also the role of 
Eqs.(\ref{4.2f}) and (\ref{4.2g}) should be exchanged together with
the role of Eqs.(\ref{4.2c}) and (\ref{4.2d}). The situation is a bit 
complicated in this case and the resulting identities are explicitly 
reproduced: 
\begin{subequations}
\begin{eqnarray}
\label{4.3a}S_{\mu}\delta \Gamma^{\mu \nu \rho \sigma}_{RAAA}(P,Q,R,S) & 
                    = & \delta \Gamma^{\nu \rho \sigma}_{RAA}(P+S,Q,R)
- \delta \Gamma^{\nu \rho \sigma}_{RAA}(P,Q+S,R), \\
\label{4.3b}S_{\mu}\delta \Gamma^{\mu \nu \rho \sigma}_{ARAA}(P,Q,R,S) & 
                    = & \delta \Gamma^{\nu \rho \sigma}_{ARA}(P+S,Q,R)
- \delta \Gamma^{\nu \rho \sigma}_{ARA}(P,Q+S,R), \\
\label{4.3c}S_{\mu}\delta \Gamma^{\mu \nu \rho \sigma}_{AARA}(P,Q,R,S) & 
                    = & \delta \Gamma^{\nu \rho \sigma}_{AAR}(P+S,Q,R)
 - \delta \Gamma^{\nu \rho \sigma}_{AAR}(P,Q+S,R), \\
\label{4.3d}S_{\mu}\delta \Gamma^{\mu \nu \rho \sigma}_{AAAR}(P,Q,R,S) & 
                    = & \delta \Gamma^{\nu \rho \sigma}_{RAA}(P+S,Q,R)
- \delta \Gamma^{\nu \rho \sigma}_{ARA}(P,Q+S,R), \\
\label{4.3e}S_{\mu}\delta \Gamma^{\mu \nu \rho \sigma}_{RRAA}(P,Q,R,S) & 
                    = & \delta \Gamma^{\nu \rho \sigma}_{RRA}(P+S,Q,R)
- \delta \Gamma^{\nu \rho \sigma}_{RRA}(P,Q+S,R), \\
\label{4.3f} S_{\mu}\delta \Gamma^{\mu \nu \rho \sigma}_{RARA}(P,Q,R,S) & 
                    = & \delta \Gamma^{\nu \rho \sigma}_{RAR}(P+S,Q,R)
- \delta \Gamma^{\nu \rho \sigma}_{RAR}(P,Q+S,R), \\
\label{4.3g}
S_{\mu}\delta \Gamma^{\mu \nu \rho \sigma}_{RAAR}(P,Q,R,S) & 
                    = &  - \delta \Gamma^{\nu \rho \sigma}_{RRA}(P,Q+S,R).
\end{eqnarray}
\end{subequations}
It is worth noting that Eqs.(\ref{4.2e}-g) and (\ref{4.3e}-g) are the 
new QED-type Ward-Takahashi identities satisfied among the vertices 
with the $O(g^2T^3)$ behavior thus being absent in the ITF.

\subsection{Ward-Takahashi identities between the two- and three-point 
vertex functions in QCD/QED}

\subsubsection{Between the 2- and 3-point HTL vertex functions with 
a fermion pair legs}
Since only the difference between in QCD and in QED is the group
factor, by comparing the results Eqs.(III.11) and Eqs.(III.6)  it is 
easy to see that we can verify the same Ward-Takahashi identities in 
QCD and in QED between vertices shown in Fig.\ref{fig3} 
(i.e.,\ref{fig4}) and Fig.\ref{fig1}:
\begin{subequations}
\begin{eqnarray}
R_{\mu}\delta \Gamma^{\mu}_{RAA}(P,Q,R) & 
                   = & \delta \Sigma_{RA}(P,Q+R) - \delta \Sigma_{RA}(P+R,Q),\\
R_{\mu}\delta \Gamma^{\mu}_{ARA}(P,Q,R) &
                   = & \delta \Sigma_{AR}(P,Q+R) - \delta \Sigma_{AR}(P+R,Q),\\
R_{\mu}\delta \Gamma^{\mu}_{AAR}(P,Q,R) & 
                   = & \delta \Sigma_{AR}(P,Q+R) - \delta \Sigma_{RA}(P+R,Q),\\
R_{\mu}\delta \Gamma^{\mu}_{RRR}(P,Q,R) & 
                   = & \delta \Sigma_{RA}(P,Q+R) - \delta \Sigma_{AR}(P+R,Q).
\end{eqnarray}
\end{subequations}

\subsubsection{Between the 2- and 3-gluon HTL vertex functions}
In QED there are no 3-photon vertex functions, thus we study in QCD 
the relation between the 2- and 3-gluon HTL vertex functions. 
Comparing the results Eqs.(III.18) and Eqs.(III.8) we obtain the 
following Ward-Takahashi identities between vertices shown in 
Fig.\ref{fig5} (i.e., \ref{fig6}) and Fig.\ref{fig2}:
\begin{subequations}
\begin{eqnarray}
\label{4.5a}
R_{\mu}\delta \Gamma^{\mu \nu \rho}_{RAA}(P,Q,R) & 
  = & \delta \Pi^{\nu \rho}_{RA}(P,Q+R) - \delta \Pi^{\nu \rho}_{RA}(P+R,Q),\\
\label{4.5b}
R_{\mu}\delta \Gamma^{\mu \nu \rho}_{ARA}(P,Q,R) & 
= & \delta \Pi^{\nu \rho}_{AR}(P,Q+R) - \delta \Pi^{\nu \rho}_{AR}(P+R,Q),\\
\label{4.5c}
R_{\mu}\delta \Gamma^{\mu \nu \rho}_{AAR}(P,Q,R) & 
   = & \delta \Pi^{\nu \rho}_{AR}(P,Q+R) - \delta \Pi^{\nu \rho}_{RA}(P+R,Q),\\
\label{4.5d}
R_{\mu}\delta \Gamma^{\mu \nu \rho}_{RRA}(P,Q,R) & 
   = & \delta \Pi^{\nu \rho}_{RR}(P,Q+R) - \delta \Pi^{\nu \rho}_{RR}(P+R,Q),\\
\label{4.5e}
R_{\mu}\delta \Gamma^{\mu \nu \rho}_{RAR}(P,Q,R) & 
                 = & \delta \Pi^{\nu \rho}_{RR}(P,Q+R), \\
\label{4.5f}
R_{\mu}\delta \Gamma^{\mu \nu \rho}_{ARR}(P,Q,R) & 
                 = &  - \delta \Pi^{\nu \rho}_{RR}(P+R,Q),\\
\label{4.5g}   
R_{\mu}\delta \Gamma^{\mu \nu \rho}_{RRR}(P,Q,R) & 
                 = & \delta \Pi^{\nu \rho}_{RA}(P,Q+R) - \delta 
                    \Pi^{\nu \rho}_{AR}(P+R,Q).
\end{eqnarray}
\end{subequations}
Here again we should note that Eqs.(\ref{4.5e}-f) are the new 
QED-type Ward-Takahashi identities satisfied among vertex functions 
with the $O(g^2T^3)$ behavior thus being absent in any amplitude in the ITF. 

In the last section \ref{s3}  we showed that in the physical
representation in the real time CPT formalism there exist those vertex
functions with two (presumably even number  in the arbitrary $n$-point
case) external retarded indices having the high temperature behavior of
$O(g^2T^3)$. In the ITF there are no $n$-point functions with the
$O(g^2T^3)$ behavior, namely all the HTL's are of $O(g^2T^2$) \cite{1,2}
among which the QED-type Ward-Takahashi identities are satisfied,
guaranteeing the gauge invariance of the HTL's. In this section we
verify that those vertex functions with the high temperature behavior of
$O(g^2T^3)$ also satisfy among themselves the simple QED-type
Ward-Takahashi identities in the HTL approximation, thus can show
explicitly the gauge invariance of  the HTL's in the real time thermal
QCD/QED. 

\vspace{0.3cm}
To close this section it is better to make notice on the identities 
(\ref{4.2f}) and (\ref{4.3g}), and also on those (\ref{4.5e}) and 
(\ref{4.5f}). They seem to have a bit different structure from others; 
identities  (\ref{4.2f}), (\ref{4.3g}), (\ref{4.5e}) and (\ref{4.5f}) 
have the right-hand-sides with a single HTL vertex functions, while 
the right-hand-sides of other identities in Eqs.(IV.2), (IV.3) and 
(IV.5) consist of difference of two vertex functions, being familiar 
in the Ward-Takahashi identities in QED.  We should note,  however, 
that the right-hand-sides of Eqs.(\ref{4.2f}), (\ref{4.3g}), 
(\ref{4.5e}) and (\ref{4.5f}) are the HTL contributions to the 3-, 
or 2-gluon vertex functions with two retarded indices, which are 
essentially consist of the difference of two vertex functions. With 
this fact we can understand that all the identities (\ref{4.2a}-g),
(\ref{4.3a}-g) and (\ref{4.5a}-g) actually have the same structure. Further 
discussion on this point will be given in the last section \ref{s6}.

\section{Dyson-Schwinger equation in the HTL approximation}
\label{s5}

Having determined all possible ingredients, namely the HTL
contributions to the 2-, 3-, and 4-point vertex functions,  now we can 
write down the Dyson-Schwinger (DS) equations in the HTL
approximation. Before doing this, however, let us here remember the 
reason why we have calculated the HTL vertex functions in the physical 
representation in the real time CTP formalism of thermal field theory. 

As we have already noticed in section \ref{s3}, to investigate the
consequences on the physical mass we need to study the mass function of
the inverse of the retarded propagator. To investigate the chiral phase
transition we need the DS equation for the retarded component of the
fermion mass function, $\Sigma_R$, and to study the magnetic screening
the DS equation for the retarded component of the gluon polarization
tensor, $\Pi^{\mu \nu}_R$. Nevertheless in many analyses \cite{13} the
\{11\}-component of the fermion self energy in the single-time
representation, $\Sigma_{11}$, not the $\Sigma_R$, have been studied by
neglecting its imaginary part without much attention. As is well known,
$Re\Sigma_R = Re\Sigma_{11}$,  but not for their imaginary parts. The
lesson from the HTL resummed perturbation theory \cite{14,15} remainds
us of the fact that the imaginary parts of  $\Sigma_R$ and of 
$\Pi^{\mu \nu}_R$ really get the important thermal effects, which in 
turn affecting their real parts. In this sense it is important to 
correctly construct the DS equations exactly for  $\Sigma_R$ and for 
$\Pi^{\mu \nu}_R$.

\vspace{0.3cm}
Now we are ready to write down the DS equations. For definiteness in
this paper we only give the  DS equation for the fermion self-energy
$\Sigma_R$ in the HTL approximation, that can be obtained by applying
the following approximation to the full DS equation;

  i) replace the full gauge boson propagator with the HTL resummed
  propagator, and 

 ii) approximate the full vertex functions to the HTL resummed vertex 
  functions.   

Then we get in QCD the desired DS equation (in case of QED,  $g^2C_F$ 
in Eq.(\ref{5.1}) should read $e^2$ and  $m_g^2$ in Eqs.(V.3) should 
be properly understood),
\begin{eqnarray}
- i\Sigma_R (P) & = &  - g^2\frac{C_F}{2} \int \frac{d^4K}{(2\pi)^4} 
       \nonumber \\
& \times &              \left\{ ^*\Gamma^{\mu}_{RAA}(-P,K,P-K) S_{RA}(K) 
                  \ ^ *\Gamma^{\nu}_{RAA}(-K,P,K-P) \right.\nonumber \\
& & \hspace{1cm} \times
\frac{2T}{p_0 - k_0}[\ ^*G_{RA, \mu\nu}(P-K)    - \ ^ *G_{AR, \mu\nu}(P-K)] 
\nonumber \\
& + &  
  \ ^*\Gamma^{\mu}_{RAA}(-P,K,P-K) (1-2n_F(k_0))\mbox{sgn}(k_0)
[S_{RA}(K) -  S_{AR}(K)] \nonumber \\
\label{5.1} & &  \hspace{3.5cm} \left. \times
\ ^*\Gamma^{\nu}_{AAR}(-K,P,K-P) \ ^*G_{RA, \mu\nu}(P-K) \right\},
\end{eqnarray}
where $\ ^*\Gamma^{\mu}$ is the HTL resummed quark-gluon vertex, defined
in Eqs.(III.9)  and given in Eqs.(III.10) with $ \gamma^{\mu}_{RAA} =
\gamma^{\mu}_{AAR} = \gamma^{\mu}$.  $\ ^*G^{\mu\nu}$ is the HTL
resummed gluon propagator, where retarded/advanced propagator  is given by
\begin{subequations}
\begin{eqnarray}
\label{5.2a}\ ^ *G^{\mu\nu}_{RA/AR}(-K,K) &=&  
 \frac1{\ ^*\Pi^{R/A}_T(K) -K^2 \mp i \mbox{sgn}(k_0)\varepsilon} 
A^{\mu\nu} 
+\frac1{\ ^*\Pi^{R/A}_L(K) - K^2 \mp i \mbox{sgn}(k_0)\varepsilon} B^{\mu\nu}
\nonumber \\
& &-\frac{\xi}{K^2 \mp i \mbox{sgn}(k_0)\varepsilon} D^{\mu\nu} ,
\end{eqnarray}
where is the gauge fixing-parameter ($ \xi=0 $ in the Landau gauge) and
the $\{RR\}$-component by
\begin{equation}
\label{5.2b} \ ^*G^{\mu\nu}_{RR}(-K,K)  = 
\frac{2T}{k_0}   
\{\ ^*G^{\mu\nu}_{RA}(-K,K)- \ ^*G^{\mu\nu}_{AR}(-K,K)\}\ ,
\end{equation}
\end{subequations}
with $\ ^*\Pi^{R/A}_T$ and  $\ ^*\Pi^{R/A}_L$ being the HTL contributions to
the retarded/advanced gluon self-energy of the  transverse and
longitudinal  modes, respectively,
\begin{subequations}
\begin{eqnarray}
\ ^*\Pi^{R/A}_T(K) & = &  m_g^2 \frac{k_0}{2k^2} 
\left(k_0+\frac{k_0^2-k^2}{2k} \log \frac{k_0-k \pm i \varepsilon}
{k_0+k \pm i \varepsilon}\right),\\
\ ^*\Pi^{R/A}_L(K) & = & -  m_g^2\frac{ k_0^2 -  k^2}{k^2} 
\left(1+\frac{k_0}{2k}\log \frac{k_0-k \pm i \varepsilon}{k_0+k \pm i 
\varepsilon}\right),
\end{eqnarray}
\end{subequations}
where $m_g$ denotes the thermal gluon
mass (in QED, $g^2(N_c+N_f/2)$ should read $e^2$),
\[ m_g^2 = \frac13g^2T^2 \left( N_c+\frac12N_f \right). \]
In Eq.(\ref{5.2a}),  $A^{\mu\nu}, B^{\mu\nu}$ and $D^{\mu\nu}$ are the 
projection tensors \cite{16}
\begin{subequations}
\begin{eqnarray}
A^{\mu\nu}(K) & \equiv &   g^{\mu\nu}-  B^{\mu\nu}(K)- D^{\mu\nu}(K),\\
B^{\mu\nu}(K) & \equiv &   - \frac{\tilde{K}^{\mu} \tilde{K}^{\nu}}{K^2},\\
D^{\mu\nu}(K) & \equiv &  \frac{K^{\mu} K^{\nu}}{K^2},\\
\tilde{K}^{\mu} & \equiv &  (k, k_0\hat{\mbox{\boldmath{$k$}}}), 
     \ \ \ \  \tilde{K}^2 = -K^2=- (k_0^2- k^2),
\end{eqnarray}
\end{subequations}
where $\hat{\mbox{\boldmath{$k$}}} \equiv {\mbox{\boldmath{$k$}}}/k$
is the unit three vector along the direction of {\mbox{\boldmath{$k$}}}.

\begin{figure}
\includegraphics{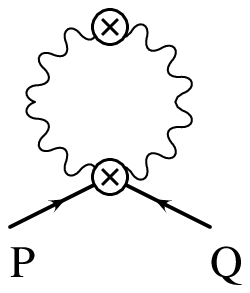}
\caption{\label{fig11}Possible contribution from the
   fermion-pair-2-gauge boson HTL vertex function to the HTL resummed
   DS equation for the fermion self-energy function $\Sigma_R$.
   $\bigotimes$ denotes the corresponding vertex function and
   propagator being the HTL resummed ones.}
\end{figure}

$S_{RA/AR} $ is the retarded/advanced full fermion propagator,
\begin{subequations}
\begin{equation}
S_{RA/AR}(-P,P)  = \frac1{\not{\!P} \pm  i\varepsilon \gamma^0- \Sigma_{R/A}},
\end{equation}
and $S_{RR}$ the $\{RR\}$-component or the correlation,
\begin{equation}
   S_{RR}(-P,P) = (1-  2n_F(p_0)) \mbox{sgn}(p_0)
\{S_{RA}(-P,P) -  S_{AR}(-P,P)\}.
\end{equation}
\end{subequations}
In the present HTL approximation, the DS equation (\ref{5.1}) becomes an
integral  equation for the unknown fermion self-energy function
$\Sigma_R$.  It is worth giving some comments on this DS equation,
(\ref{5.1}):

  i) The HTL resummed quark-gluon vertex function is substituted for
  both of vertices in Eq.(\ref{5.1}). This may cause the double counting
  problem when the loop-momentum becomes "hard". Thus in the actual
  analysis we should introduce an intermediate momentum scale to cut the
  loop-momentum integration. In the hard loop-momentum region the naive
  ladder approximation may work, while  in the soft loop-momentum region
  we must use Eq.(\ref{5.1}).

 ii) There are no contributions from the fermion-pair-2-gluon vertex
 function in Eq.(\ref{5.1}).  In QCD this happens quite miraculously,
 namely the contribution to $\Sigma_R$ from the diagrams shown in 
 Fig.\ref{fig11} are totally and exactly accounted for through the 
 loop diagram with two
 3-point quark-gluon vertices in Eq.(\ref{5.1}). Inclusion of the
 contribution from Fig.\ref{fig11} causes the trouble of double counting
 of diagrams.  

 iii) The first term in the curly bracket, which exists also in the 
limit of ladder approximation,  has been dismissed in previous DS 
equation analyses \cite{13}. As can be seen this term could produce 
an dominant contribution due to the presence of the enhanced 
temperature dependence. Result of the analysis of Eq.(\ref{5.1}) will 
be given elsewhere \cite{17}.

\vspace{0.3cm} 
Similarly we can write down the DS equation for the gauge boson
polarization tensor, with which the full QCD analysis may be performed.
In the DS equation for the gluon polarization tensor, which is not given
here explicitly, the contribution from the gluon 4-point vertex does
exist. However, in this case also contributions from the HTL's, $\delta
\Gamma^{\mu \nu \rho \sigma}$, Eqs.(III.24), are totally and exactly
accounted for through the loop diagram with two 3-point gluon vertices,
thus to avoid the double counting problem only the contribution from the
tree 3-gluon vertex survives. Results of the analysis including the DS
equation for the gauge boson polarization tensor itself will also be
presented elsewhere \cite{17}.   

\section{Conclusions and Discussion}
\label{s6}

In this paper we calculated the $n$-point HTL vertex functions in QCD/QED
for  $n$= 2, 3 and 4 in the physical representation in the RTF. The 
result showed that the $n$-point HTL vertex functions can be
classified into two groups,  a) those with odd numbers of external 
retarded indices,  and b) the others with even numbers of external 
retarded indices.  The $n$-point HTL vertex functions with one
retarded index, which obviously belong to the first group a), are 
nothing but the HTL vertex functions that appear in the ITF, and vise
versa \cite{6}. All the HTL vertex functions belonging to the first 
group a) are of  $O(g^2T^2)$ ,  and satisfy among them the simple 
QED-type Ward-Takahashi identities, as in the ITF \cite{1,2}.  All 
the HTL vertex functions with a fermion pair belong to this first 
group a). Those vertex functions belonging to the second group b) 
never appear in the ITF, namely their existence is characteristic of 
the RTF,  and their HTL's have the high temperature behavior of 
$O(g^2T^3)$, one-power of $T$ higher than usual. Despite of this 
difference we could verify that those HTL vertex functions belonging 
to the second group b) also satisfy among themselves the QED-type 
Ward-Takahashi identities, thus guaranteeing the gauge invariance of 
the HTL's in the real time thermal  QCD/QED. The group b) HTL's
consist of the $n$-gluon vertex functions.

\vspace{0.3cm}
\noindent
Comments  and discussion are in order.

  i) With the present results we can feel at ease to perform in the 
framework of real time thermal field theories the nonperturbative 
analyses in the HTL approximation on dynamical phenomena such as the 
chiral phase transition.  As an application we derived the HTL
resummed DS equation for the retarded fermion self-energy function, 
enabling us to study in QCD/QED the chiral phase transition at finite 
temperature, which is now under investigation \cite{17}.

  ii) It is well known that all the HTL contribution in the ITF are 
of order  $O(g^2 T^2)$.  Are there anything wrong in our  results, 
Eqs. (\ref{3.8c}), (\ref{3.18e}-g), and (\ref{3.24e}-g), contradicting 
to the previous results? The answer is NO.  It is shown \cite{6} that 
the $n$-point functions in the ITF is nothing but the special 
combination of those  in the RTF, namely, just the $n$-point 
functions in the real time physical representation with only one 
external retarded index, all the other ($n$-1) external indices being 
the advanced ones. This means that the $n$-point functions with more 
than two external retarded indices never have their counterparts in 
the ITF. Furthermore we have already had some results indicating the 
existence of the order  $O(g^2 T^3)$ contributions in the RTF 
\cite{8,9}. Our results as above have clearly shown that such order  
$O(g^2 T^3)$ contributions really exist in the real time physical 
representation.  Of course these unusual HTL's appear only in the 
$n$-gluon vertex functions in QCD. In QED there are  no 3-photon 
vertex functions and no 4-photon HTL vertex functions, as already shown.

  iii) Another point to be noted is that the results, say, 
Eqs. (\ref{3.8a}-c) have nothing in trouble with the number of 
independent components which is in the present case only one. 
As we can easily see the HTL contributions satisfy Eq.(\ref{3.2a}),
\[
       \delta \Pi^{\mu \nu}_{AR}(P,Q) = (\delta \Pi^{\mu
       \nu}_{RA}(P,Q))^*.
\]
Remembering that Eq.(\ref{3.2b}) holds among the full two point vertex 
functions and here we are studying the HTL contributions, we can see 
that the additional power of $T$ in Eq.(\ref{3.6c}) comes from the 
boson distribution function, guaranteeing the number of independent 
components in the present case being only one. The same is true for 3- 
and 4-point vertex functions with more than two external retarded indices.

 iv) As noticed at the end of section \ref{s4}, although the 
Ward-Takahashi identities (\ref{4.2f}) and (\ref{4.3g}), and also 
those (\ref{4.5e}) and (\ref{4.5f}) seem to have a bit different 
structure from others in Eqs.(IV.2), (IV.3) and (IV.5), all the 
identities (\ref{4.2a}-g), (\ref{4.3a}-g) and (\ref{4.5a}-g) actually 
have the same structure. This fact can be clearly seen by noticing 
that, e.g., the right-hand-sides of Eqs. (\ref{4.5e}) and (\ref{4.5f}) 
are the HTL contributions to the $\{RR\}$-component of gluon 
polarization tensor which is actually the difference of two 
polarization tensors,
\begin{equation}
\label{6.1}
\Pi^{\mu \nu}_{RR}(P,Q) =
         (1- 2n_B(q_0))\mbox{sgn}(q_0)
( \Pi^{\mu \nu}_{RA}(P,Q) - \Pi^{\mu \nu}_{AR}(P,Q) ),  \ \ P+Q=0, 
\end{equation}
namely, in the HTL approximation (where the external momentum $-P=Q$ 
must be soft)
\begin{equation}
\label{6.2}
\delta\Pi^{\mu \nu}_{RR}(P,Q) \hat{=}
        \frac{2T}{q_0}( \delta\Pi^{\mu \nu}_{RA}(P,Q) - \delta
        \Pi^{\mu \nu}_{AR}(P,Q) ).  
\end{equation}
Eq.(\ref{6.2}) shows that the right-hand-sides of Eqs. (\ref{4.5e}) 
and (\ref{4.5f}) are
nothing but the difference of two HTL gluon polarization tensors, thus
having exactly the same structure as others. The same is true for the
right-hand-sides of identities (\ref{4.2f}) and (\ref{4.3g}), though 
less clear but can be easily seen with explicit manipulations.

   v) We showed explicitly that in the physical  representation in 
the RTF there exist those vertex functions with two external 
retarded indices having the high temperature behavior of $O(g^2T^3)$. 
This  should be true for any $n$-gluon vertex functions with even 
number external retarded indices.

\begin{acknowledgments}
 We thank the useful discussion at the Workshop on Thermal Quantum Field
Theories and their Applications, held at the Yukawa Institute for
Theoretical Physics, Kyoto, Japan, 6 -- 8 August, 2001. This work is
partly supported by Grant-in-Aid of Nara University, 2001 (HN).
\end{acknowledgments}


\newpage

\vspace{1cm}

\vspace{1cm}

\vspace{1cm}

\vspace{1cm}

\vspace{1cm}

\vspace{1cm}

\vspace{1cm}

\vspace{1cm}

\vspace{1cm}

\vspace{1cm}


\begin{thebibliography}{99}
\bibitem{f2}
  Our notations are the same as those used in Ref.\cite{7}, being 
  slightly different from the originalones in Ref.\cite{4}. Details 
  of the definitions and notations are given in section \ref{s2}.
\bibitem{f3}
  This result differs from the one in Ref.\cite{7}, in which it is 
  claimed that there exist the non-vanishing HTL contributions in 
  the 2f-2g 4-point vertex functions with two retarded indices, being
  proportional to $g^2T^2$.  The corresponding Ward-Takahashi
  identities they have claimed to hold are nothing but ``0=0'' in our case.
\end{thebibliography}
\end{document}